\documentstyle[prl,aps]{revtex}

\def\Ac{\mbox{{$\cal A$}}}

\draft
\begin{document}
\input{epsf}
\twocolumn[\hsize\textwidth\columnwidth\hsize\csname@twocolumnfalse\endcsname

\title{Multi-valued mappings in generalized chaos synchronization}

\vspace{0.4in}

\author{ N.F. Rulkov$^{a}$, V.S. Afraimovich$^{b}$, C.T. Lewis$^{a}$,
J.-R. Chazottes$^{b,d}$ and A. Cordonet$^{b,c}$}
\vspace{0.1in}

\address{
 $^{a}$ Institute for Nonlinear Science, University of California,
San Diego, La Jolla, CA 92093-0402}
\address{
 $^{b}$ IICO-UASLP, A. Obreg\'on 64, 78000 San Luis Potos\'{\i},
SLP, M\'exico }
\address{
 $^{c}$ Centre de Physique Th\'eorique, Universit\'e de la
M\'editerran\'ee, Luminy Case 907, F-13288 Marseille Cedex 9,
France}
\address{
 $^{d}$ IME-USP, R. do Mat\~ao, 1010, 05508-900 S\~ao Paulo, Brazil}
\date{\today}
\maketitle

\begin{abstract}

The onset of generalized synchronization of chaos in
directionally-coupled systems corresponds to the formation of a
continuous mapping which enables one to persistently define the
state of the response system from the trajectory of the drive
system. The recently developed theory of generalized
synchronization of chaos deals only with the case where this
synchronization mapping is a single-valued function. In this paper,
we explore generalized synchronization in a regime where the
synchronization mapping can become a multi-valued function.
Specifically, we study the properties of the multi-valued
mapping which occurs between the drive and response systems
when the systems are synchronized with a frequency ratio other
than one-to-one, and address the issues of the existence and
continuity of such mappings. The basic theoretical framework
underlying the considered synchronization regimes is then
developed.

\end{abstract}
\tableofcontents
\narrowtext
\vskip1pc]

\section{Introduction}\label{introduction}
The synchronization of oscillations is one of the most interesting
nonlinear phenomena and is an inherent part of many processes
studied in a wide range of natural systems, including such diverse
areas as neuro-biological networks and solitary systems. As a
result, the corresponding theory is extensively utilized in many
practical applications~\cite{Minorsky,Blekhman,Glass}. The
dynamical theory of the synchronization of periodic oscillations,
which relates the onset of synchronization to the birth of a stable
limit cycle due to the bifurcation of the motion on a
two-dimensional ergodic torus, is due to Van der Pol~\cite{VDP}.
Since the discovery of chaotic behavior in nonlinear oscillators
and the ability of chaotic oscillators to synchronize, the
framework of the dynamical theory of synchronization has been
significantly modified. The theory now encompasses the major
properties inherent in the synchronization that deals with limiting
sets (called chaotic attractors) which are dynamically much richer than
isolated limit cycles~\cite{AVR86}.

Different notions of synchronization occurring in chaotic
oscillators have been introduced in order to explore the
qualitative dynamical changes caused by the onset of chaos
synchronization in numerous experimental studies. Such notions
include the cases of {\em identical} synchronization, where
identical chaotic oscillations are usually studied in coupled
systems with identical individual dynamics~\cite{FY84,Pecora90},
{\em generalized} synchronization, which extends the notion of
identical synchronization to cases of directionally-coupled systems
with non-identical individual
dynamics~\cite{AVR86,Rulkov95,Pecora95,Parlitz96,Schft96}, and {\em
phase} synchronization which usually deals with the phase locking
of the main frequencies in the spectrum of the chaotic systems
while the chaotic components of the signals remain
independent~\cite{Pikovsky97,Zaks99}. Studies of the phenomena of
generalized and phase synchronization are largely motivated by the
need for development of a theoretical framework that can be used to
achieve a better understanding of synchronization in experiments
with neuro-biological systems where complex, chaotic behavior of
neurons is very typical~\cite{Schft96,RA98,Fitzgerald99,HDIA96}.

This paper addresses issues of generalized synchronization which
arise in studies of chaos synchronization with a frequency ratio
other than one-to-one. As it is defined in~\cite{Rulkov95}, the
onset of generalized synchronization in directionally coupled
chaotic systems relates to the formation of a continuous mapping
that transforms the trajectory on the attractor of the drive system
into the trajectory of the response system. For systems with
invertible dynamics, this is equivalent to the formation of a
continuous mapping which links the current states of the systems
once they settle down on the synchronous attractor. Since the
introduction of the concept of generalized chaos synchronization,
significant progress has been made in understanding the
relationship between the properties of these synchronization
mappings and the spectrum of Lyapunov exponents which characterize
the synchronous chaotic attractor. A number of papers have analyzed
conditions which guarantee the differentiability of the
synchronization mapping, thus indicating the formation of a
differentiable invariant manifold (in the joint phase space of the
coupled systems) which contains the synchronized chaotic attractor
(see, for example~\cite{Hunt,Josic}). Simply put, this
differentiability occurs when the rate of contraction in the
direction transverse to the manifold is larger than the rate of
contraction experienced by trajectories of the chaotic attractor in
the direction tangent to the manifold. Such cases are distinguished
as belonging to a subclass of generalized synchronization which has
been named {\em differentiable generalized
synchronization}~\cite{Hunt}.

The case of a non-differentiable synchronization mapping formed in
drive-response systems has been studied recently in~\cite{Afr2000}.
In this paper, the stability of the response behavior in the whole
phase space of the response system is linked with the existence of
a mapping which maps the trajectories of the chaotic attractor in
the driving system into the trajectories of the response system
with a mapping which is a H\"older continuous function. These
results help fill in the gap in parameter space which exists in
between the synchronization regime where a smooth synchronization
manifold exists and the natural borders of the synchronization
zones. In the present paper, we continue by studying how this
theory of continuous functions can be extended to cases of chaos
synchrony with frequency ratio other than one-to-one, where the
synchronization mapping can become multi-valued.

In section~\ref{sec2}, we present an example of chaos synchronization
with frequency ratio 2:1 and study the formation of the multi-valued
mapping in detail. We do this by examining the bifurcations
responsible for the transition from the single-valued mapping to the
multi-valued mapping as the regime of generalized synchronization of
chaos changes. In section~\ref{sec3}, in order to provide a more
complete view of the synchronization mappings and their properties,
we supplement our numerical simulation results with a discussion of
some recent rigorous mathematical results not easily available to the
reader, which are related to the considered regimes of generalized
synchronization. We then review the results of the theory behind
generalized synchronization with continuous functions and
develop a similar theory for the case of multi-valued functions.

\section{Chaos synchronization with frequency ratio 2:1}\label{sec2}

\begin{figure}
\begin{center}
\leavevmode
\hbox{%
\epsfxsize=7.2cm
\epsffile{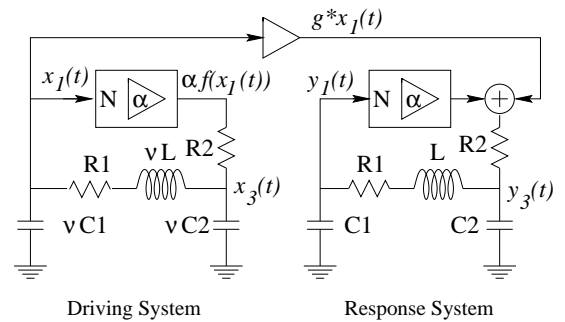}}
\end{center}
\caption{The sketch of the circuit diagram for the experimental
study of chaos synchronization with frequency ratio 2:1. The
frequency ratio is controlled by the multiplier $\nu$ used to
select the values of the inductors and capacitors in the drive and
response circuits.}
\label{fig1}
\end{figure}

We consider a regime of chaos synchronization with frequency ratio
2:1. This type of synchronization was earlier observed in the
experiments with two directionally-coupled chaotic circuits. The
sketch of the experimental setup is presented in Fig.~\ref{fig1}.
The details on the experiment and the parameters of the circuits
can be found in~\cite{Rulkov96}.

In this paper we present a detailed analysis of this
synchronization regime based on numerical simulation results.

\subsection{Model}\label{sec2a}
\begin{figure}
\begin{center}
\leavevmode
\hbox{%
\epsfxsize=6.2cm
\epsffile{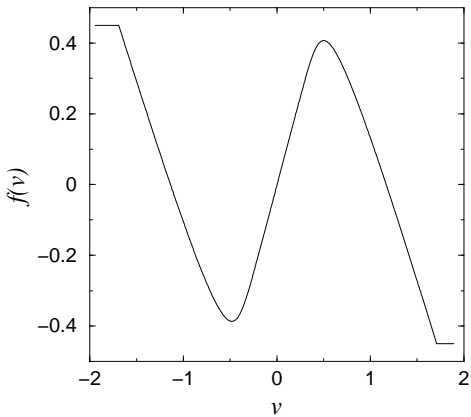}}
\end{center}
\caption{The shape of the nonlinear function $f(~)$ measured
in the experimental setup.}
\label{fig2}
\end{figure}

The dynamics of the drive circuit are described by the set of
differential equations of the form~\cite{RS97}
\begin{eqnarray}
\nu \dot{x_1} & = & x_2 \nonumber \\
\nu \dot{x_2} & = & -x_1 -\delta x_2 +x_3 \\
\nu \dot{x_3} & = & \gamma ( \alpha_1 f(x_1) - x_3 ) -\sigma x_2 \nonumber.
\end{eqnarray}
The response system equations are
\begin{eqnarray}
\dot{y_1} & = & y_2 \nonumber \\
\dot{y_2} & = & -y_1 -\delta y_2 +y_3 \\
\dot{y_3} & = & \gamma ( \alpha_2 f(y_1) - y_3 + g x_1 ) -\sigma y_2 \nonumber,
\end{eqnarray}
where $\gamma=\sqrt{\mbox{LC1}}/\mbox{R2C2}=0.294$,
$\sigma=\mbox{C1}/\mbox{C2}=11.52$,
$\delta=\mbox{R1}\sqrt{\mbox{C1}/\mbox{L}}=0.534$, $\alpha_1=15.93$
and $\alpha_2=16.7$ are the fixed system parameters and $g$ is the
coupling strength. Variables $x_1$, $x_3$, $y_1$, and $y_3$
correspond to the voltages across the capacitors (see
Fig.~\ref{fig1}). Variables $x_2$ and $y_2$ are proportional to the
current through the inductors in the drive and response circuits
respectively. The nonlinear function $f(~)$ models the input-output
characteristics of a nonlinear converter (N) used in the circuit.
The shape of the nonlinearity is presented in Fig.~\ref{fig2}. In
the numerical simulation, we modeled the function $f(x)$ with the
formulae
$$f(x)=\mbox{sign}(x)\left(a-\sqrt{d(f_p(x)-a)^2+c}\right)/d$$
where $$ f_p(x)= \cases{ |x| & if $ |x| \leq a $ \cr
-q (|x|-p)  & if $ a < |x| \leq b $ \cr
-a & if $ |x| > b $ \cr
}$$ $d=(a^2 -c)/a^2$, $q=2a/(b-a)$, $p = (b+a)/2$. The values of
the parameters $a$, $b$ and $c$ are chosen to be equal to 0.5, 1.8,
and 0.03 respectively (see~\cite{RS97} for details). The
parameter $\nu$ in the equations of the drive system is the time
scaling parameter, which is used to select the frequency ratio of
the synchronization zone.

\begin{figure}
\begin{center}
\leavevmode
\hbox{%
\epsfxsize=4.2cm
\epsffile{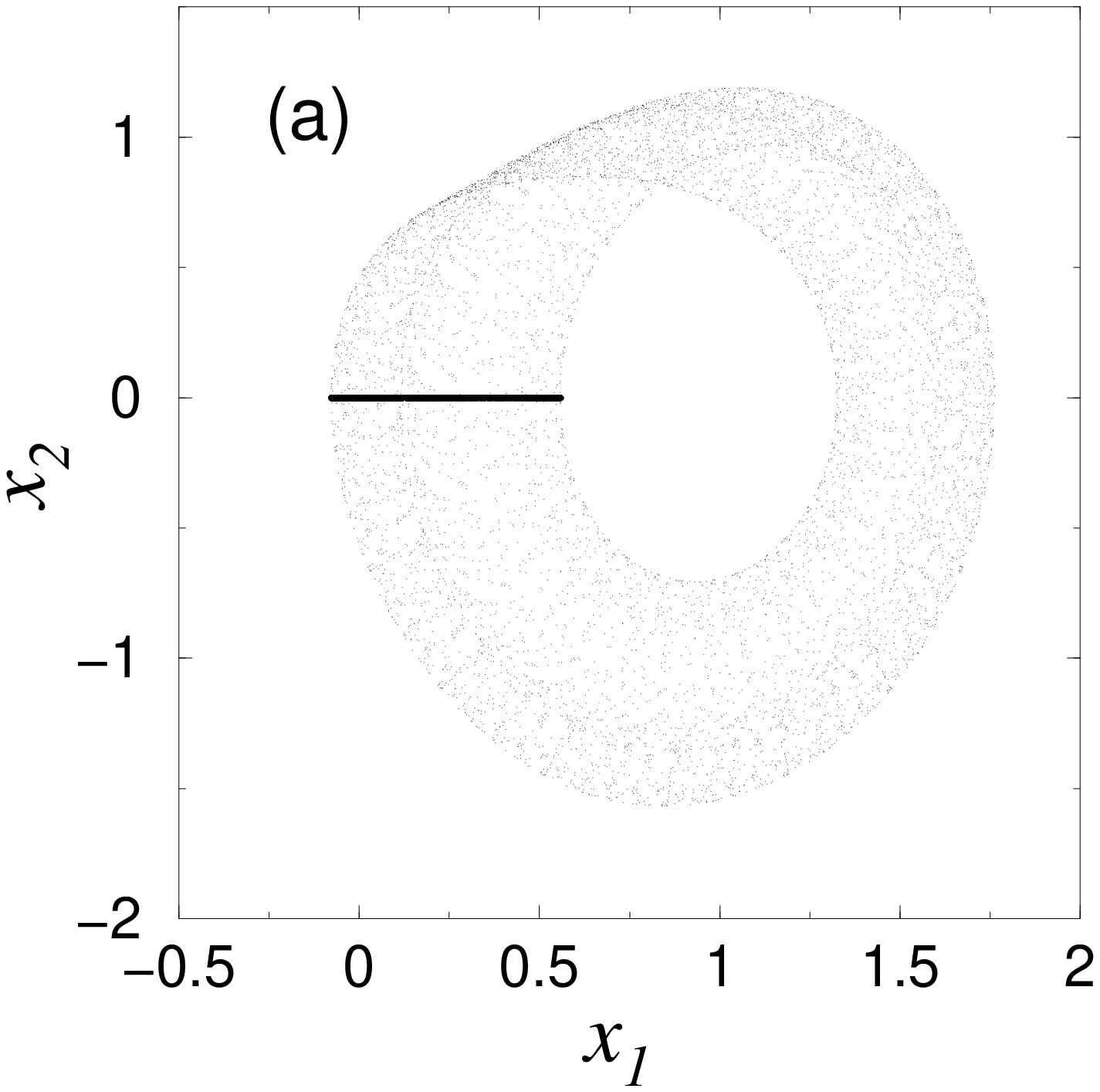}
\epsfxsize=4.2cm
\epsffile{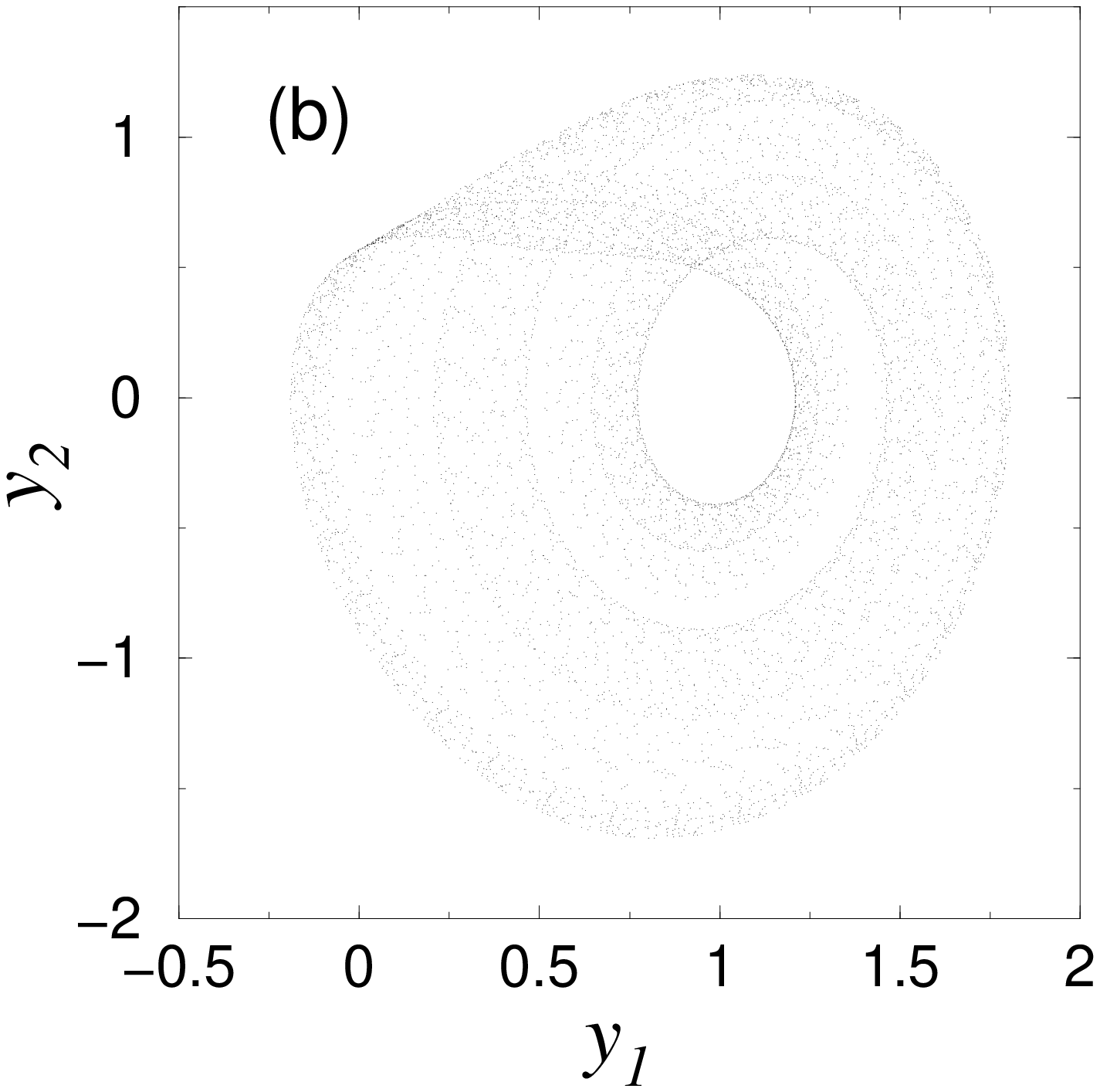}}
\end{center}
\caption{Chaotic attractors of the drive system computed with the
parameter value $\alpha_1=15.93$ (a), and the uncoupled response system
(b). The bold dots on the trajectories of the driving attractor
show the points where the chaotic trajectories cross the
Poincar\'e cross section ($x_2=0$ with $dx_2/dt>0$).}
\label{fig3}
\end{figure}

The chaotic attractors occurring in the uncoupled drive and
response systems with these parameter settings are shown in
Fig.~\ref{fig3}. Since we are interested in studying the chaos
synchronization regime with a frequency ratio of 2:1, the parameter
$\nu$ was set equal to 0.498. As a result, the phase velocity of
the trajectories of the driving attractor is about as twice as high
as that of the response attractor.

\subsection{Onset of generalized synchronization}\label{sec2b}

\begin{figure}
\begin{center}
\leavevmode
\hbox{%
\epsfxsize=4.2cm
\epsffile{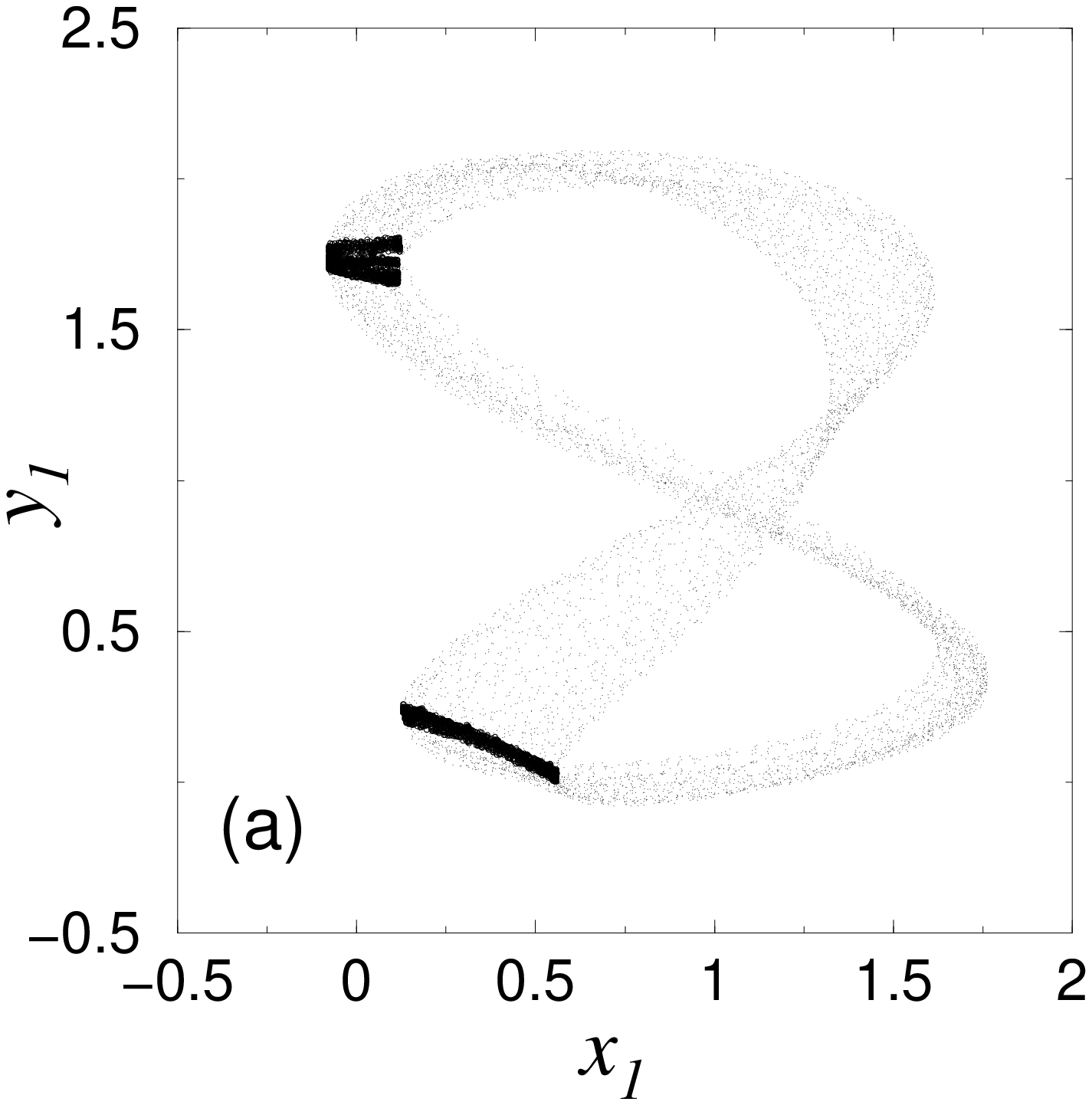}
\epsfxsize=4.2cm
\epsffile{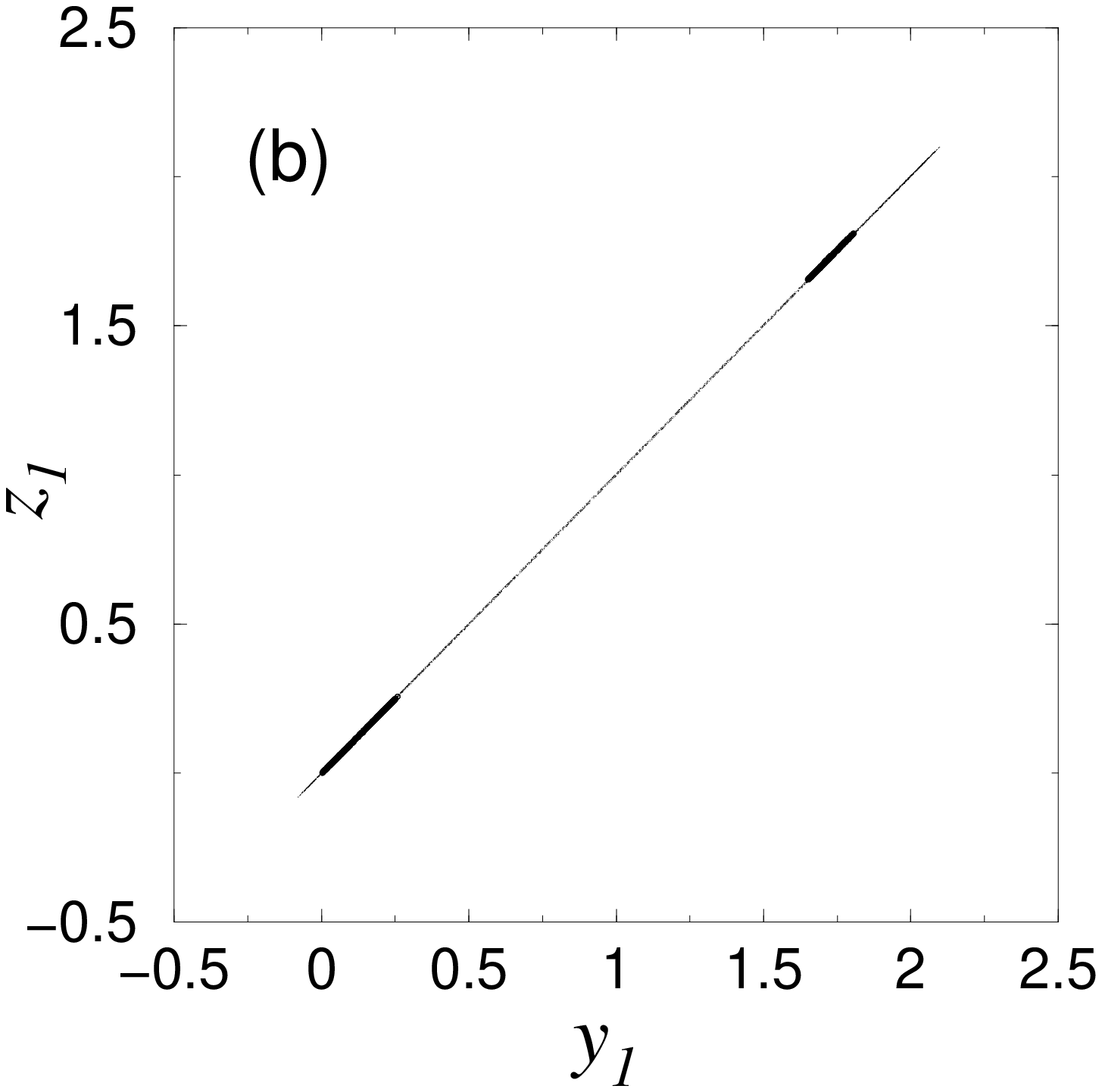}}
\end{center}
\caption{(a) The Lissajous figure for the synchronized chaotic oscillations
plotted on the plane of variables from the drive and response
($x_1,y_1$), and (b) the projections of the corresponding chaotic
trajectories onto the plane of similar variables in the response and
auxiliary systems ($y_1,z_1$). The strength of the coupling is $g=2.5$.
The bold dots show the points on the chaotic trajectories corresponding to the moments of time when the trajectory of the drive system crosses the Poincar\'e cross section ($x_2=0$) with a positive value of $dx_2/dt$ .}
\label{fig4}
\end{figure}

Introduction of sufficiently strong coupling between the systems
results in the onset of synchronization. The regime of
synchronization which occurs when $g=2.5$ is presented in
Fig.~\ref{fig4}. The onset of synchronization in this case is
detected using Lissajous figures, by observing the attractors in
the Poincar\'e cross section, and by utilizing the auxiliary system
method. Each of these methods will be briefly explained in the
following paragraphs.

The Lissajous figure is a standard approach frequently utilized in
traditional studies of synchronization between periodic
oscillators. It enables one to clearly see the onset of phase
locking and define the frequency ratio of the particular
synchronization zone. For the frequency ratio 2:1, the Lissajous
figure has a form similar to the shape of the digit ``eight".
Despite the fact that due to the chaotic behavior the Lissajous
figure does not appear as a closed curve, it is still not difficult
to see the onset of phase locking and identify the frequency ratio
(see Fig.~\ref{fig4}a).

Analysis of generalized synchronization of chaos with the help of
Poincar\'e cross sections was proposed in~\cite{Volkovskii89}. In
this analysis, we define the Poincar\'e cross section in the drive
system and detect the moments of time when driving trajectory
crosses it. At these moments, we sample the state of the response
system and compare the points to the corresponding set of the
points in the drive system. This method simplifies the analysis of
chaotic attractors by reducing their dimension by one. It helps to
detect the onset of phase locking and, in some cases, to examine
the onset of topological equivalence between the attractors in the
synchronized drive and response systems. In the numerical
simulations we analyze the properties of the chaotic attractors on
the Poincar\'e cross section ($x_2=0$) by computing the points
where the trajectories cross the value $x_2=0$ with positive values
of $dx_2/dt$ (see Fig~\ref{fig3}a).

The auxiliary system method was proposed as a practical method
of detecting the formation of a continuous mapping through the
ability of persistently point out the current state of the response
system without direct computation of the map~\cite{Abarbanel96}. This
method assumes the use of a third (auxiliary) system which is an
exact replica of the response system. In our case, the auxiliary
system is of the form
\begin{eqnarray}
\dot{z_1} & = & z_2 \nonumber \\
\dot{z_2} & = & -z_1 -\delta z_2 +z_3 \\
\dot{z_3} & = & \gamma ( \alpha_2 f(z_1) - z_3 + g x_1 ) -\sigma z_2 \nonumber.
\end{eqnarray}
Due to the identity of the dynamics of the response and auxiliary
systems, the nine dimensional space encompassing all three coupled
systems has an invariant manifold ${\bf y=z}$, where ${\bf y}$ and
${\bf z}$ are vectors of the variables in the response and
auxiliary systems, respectively. When this manifold contains the
entire chaotic attractor, a continuous mapping exists which
projects the trajectories of the drive system onto the trajectories
of the response systems after transients die out
(see~\cite{Abarbanel96} for details). Therefore, the formation of a
chaotic attractor in the invariant manifold (${\bf y=z}$)
constitutes the onset of generalized synchronization between the
drive and response systems. Such a chaotic attractor occurs in our
simulation with $g=2.5$ (see Fig~\ref{fig4}b).

When one utilizes this method in numerical simulations,
it can be difficult to ensure that the chaotic set of
trajectories in the invariant manifold does not contain
transversely unstable limit sets. Due to the identity of the
response and auxiliary systems and the limited accuracy of
numerical simulations, the chaotic trajectory can settle down on
the invariant manifold ${\bf y=z}$, even if the chaotic set in the
manifold contains transversely unstable limiting subsets. This
artifact of numerical simulation can be resolved by breaking the
symmetry between the response and auxiliary systems. In our
simulations, we accomplish this by using slightly different values of
coupling. The coupling of the auxiliary system is larger than the
value of coupling in the response system by a value of $\Delta
g=0.01$, less than 1\% of the coupling strength used.

\subsection{Two regimes of generalized synchronization}\label{sec2bnew}

\begin{figure}
\begin{center}
\leavevmode
\hbox{%
\epsfxsize=7.2cm
\epsffile{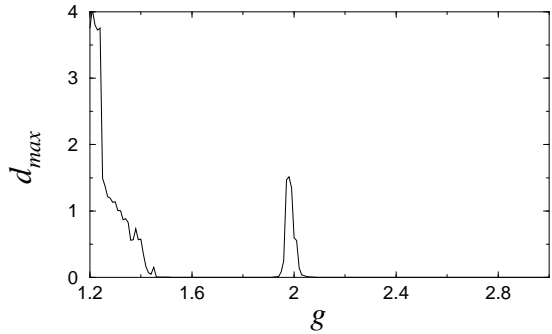}}
\end{center}
\caption{The maximal deviation between the states of the response and auxiliary
systems versus the value of coupling parameter $g$ computed with
$\Delta g=0.01$.}
\label{fig5}
\end{figure}

The goal of this section is to examine in detail the main features
of two qualitatively distinct regimes of generalized
synchronization observed for different values of coupling in some
previous experimental studies of chaos synchronization with a
frequency ratio of 2:1 (see~\cite{Rulkov96} for details on the
experimental setup). In order to point out the change between the
regimes of generalized synchronization versus variation of the
coupling parameter, we analyzed the $g$ dependence of
$d_{max}(g)=\{\sqrt{{\bf (z-y)}^T{\bf (z-y)}}\}_{max}$, which is
the maximal deviation between the response and auxiliary systems
computed for the chaotic trajectories after transients die out.
This dependence is presented in Fig.~\ref{fig5}. Synchronization
takes place for values of $g$ where $d_{max}(g)$ is approximately
equal to zero. In Fig.~\ref{fig5} there are two intervals of $g$
which correspond to two synchronization regimes. These intervals
are separated by a region of asynchronous behavior appearing around
$g=2.0$.

\begin{figure}
\begin{center}
\leavevmode
\hbox{%
\epsfxsize=4.2cm
\epsffile{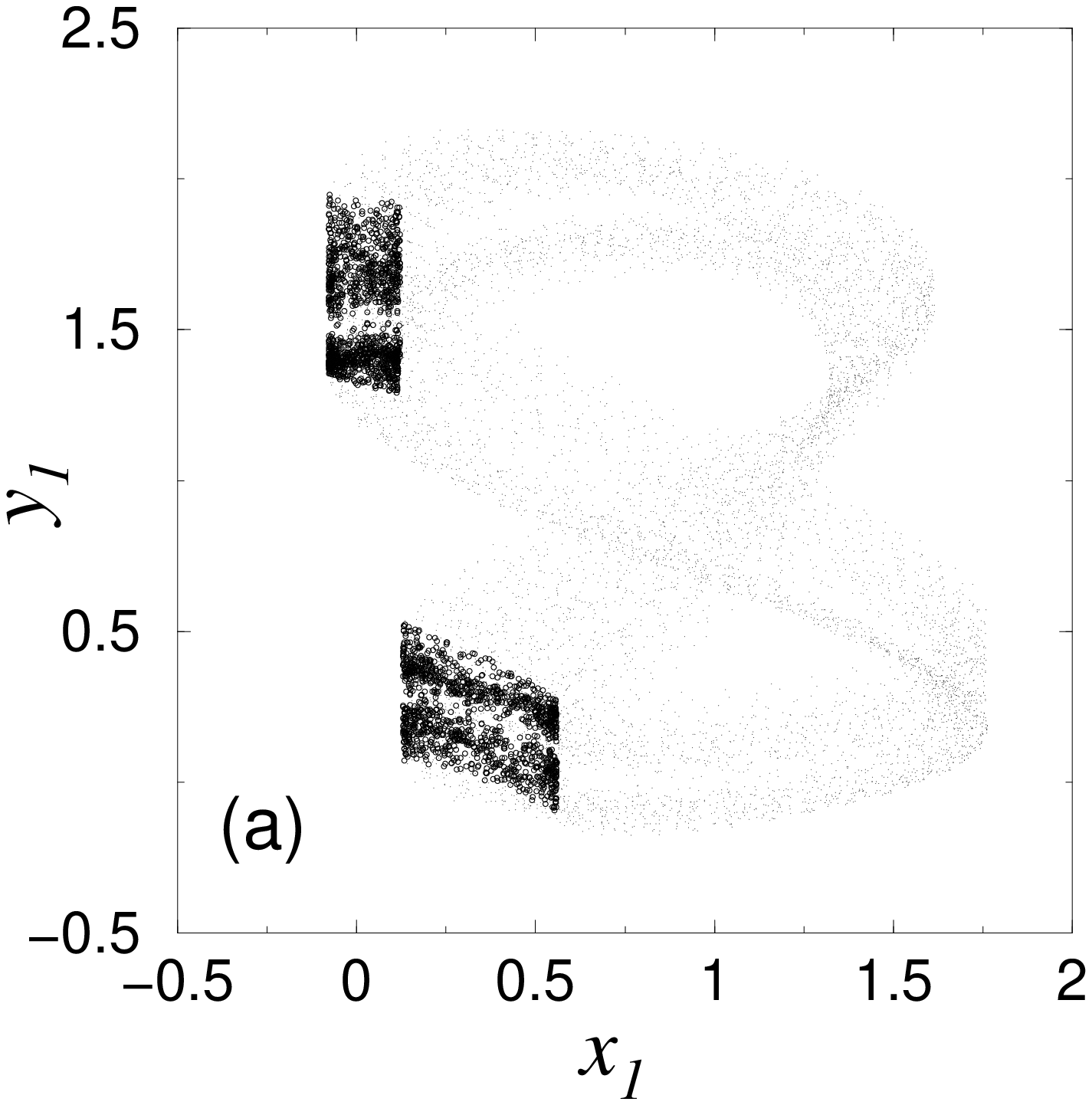}
\epsfxsize=4.2cm
\epsffile{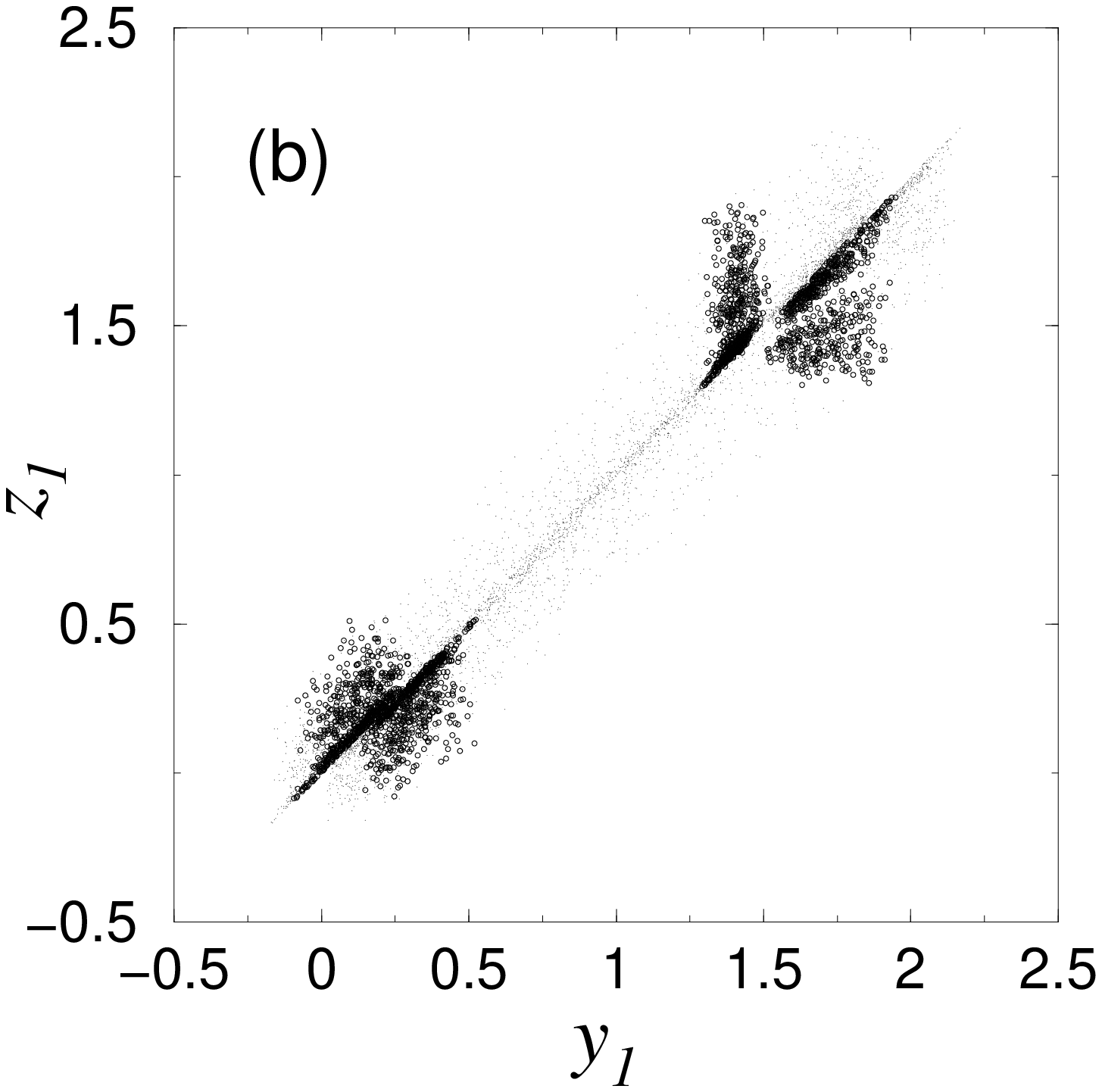}}
\end{center}
\caption{The regime of asynchronous chaotic oscillations
computed for $g=2.0$.(a) The Lissajous figure in the plane
($x_1,y_1$), and (b) the projections of the corresponding chaotic
trajectories onto the plane ($y_1,z_1$). The bold dots show the
points on the chaotic trajectories corresponding to the moments of
time when the trajectory of the drive system crosses the Poincar\'e
cross section ($x_2=0$) with a positive value of $dx_2/dt$.}
\label{fig6}
\end{figure}

The oscillations occurring in the synchronization interval $g >
2.1$ correspond to the chaotic attractor shown in Fig.~\ref{fig4}.
When the value of coupling decreases and arrives at the border of
this synchronization interval, generalized synchronization
terminates. The chaotic attractor corresponding to these
asynchronous oscillations, computed for $g=2.0$, is presented in
Fig.~\ref{fig6}. The destruction of the generalized synchronization
follows from the appearance of sporadic outbursts of non-identical
behavior in the response and auxiliary systems, which are seen in
the Fig.~\ref{fig6}b. Such outbursts are indicative of the
existence of transversely unstable limiting trajectories in the
chaotic set located in the manifold ${\bf y}={\bf z}$ and
therefore, indicates the inclusion of conditionally unstable
limiting trajectories in the chaotic attractor residing in the
phase space of the drive and response systems.

\begin{figure}
\begin{center}
\leavevmode
\hbox{%
\epsfxsize=4.2cm
\epsffile{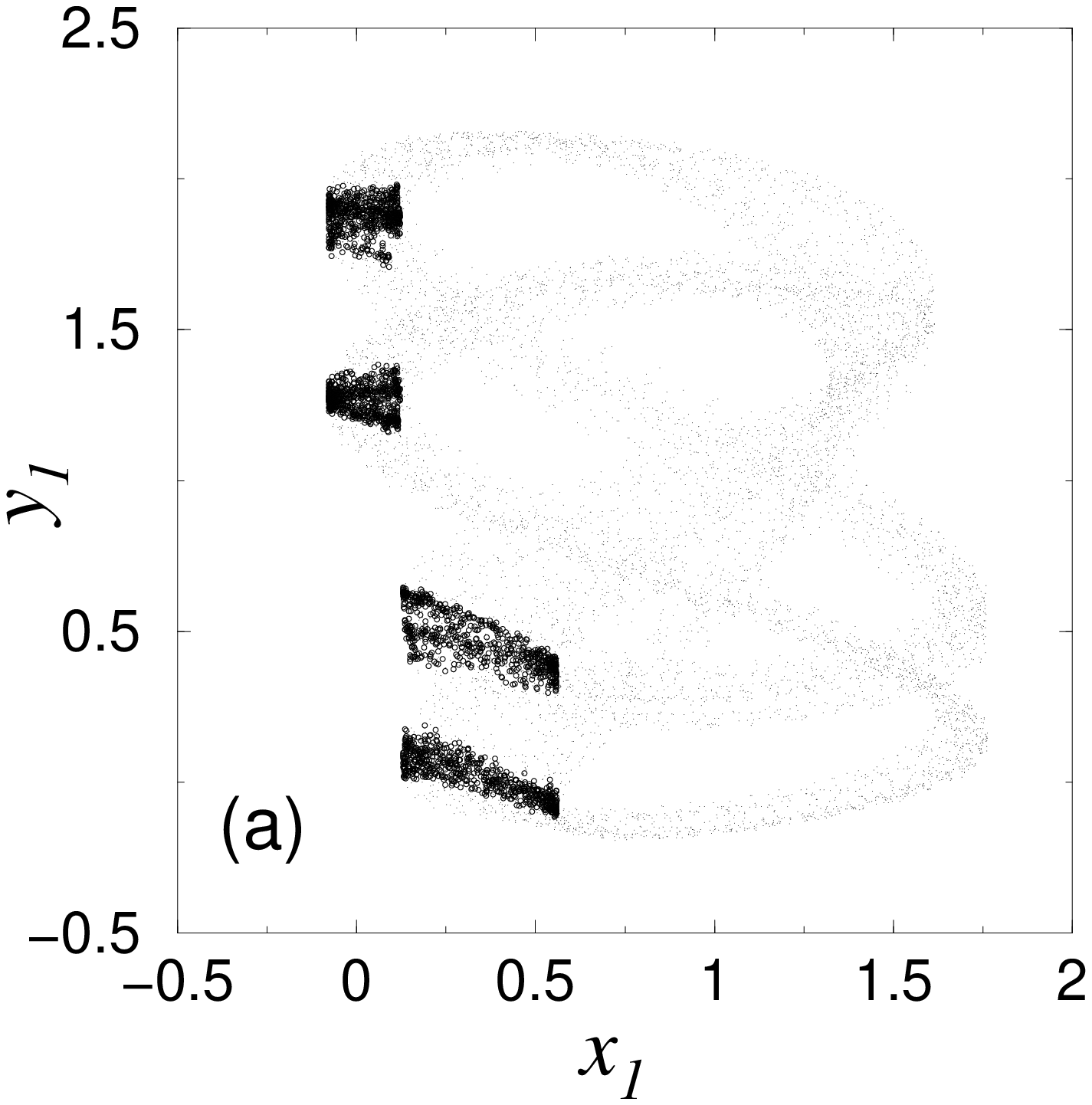}
\epsfxsize=4.2cm
\epsffile{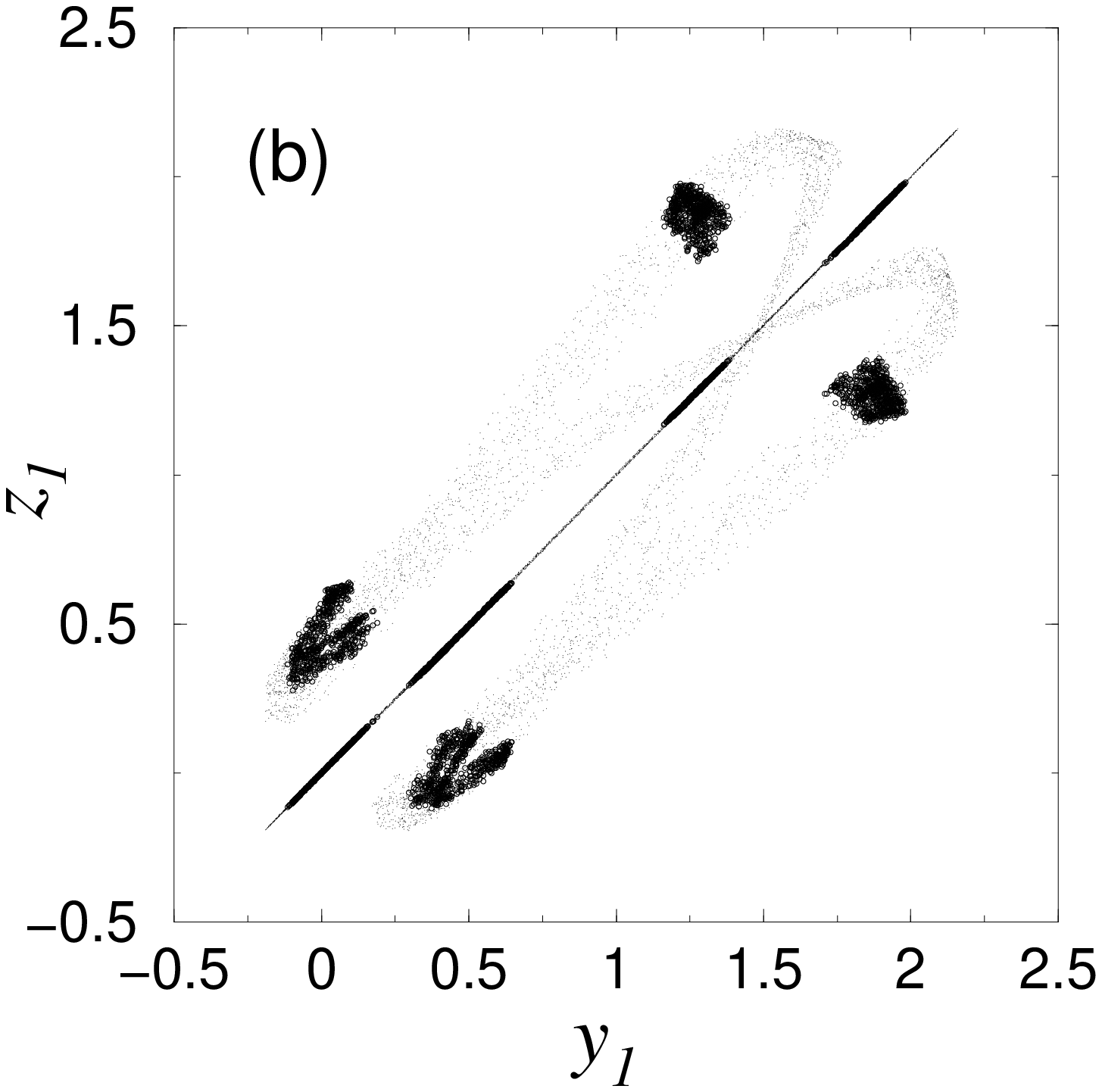}}
\end{center}
\caption{The regime of synchronized chaotic oscillations computed for
$g=1.8$. (a) The Lissajous figure of the synchronized chaotic
attractor in the plane of variables from the drive and response
systems($x_1,y_1$). (b) Two chaotic attractors corresponding to the
synchronized chaos formed in the response and auxiliary systems
plotted in the plane ($y_1,z_1$). The bold dots show the points on
the chaotic trajectories corresponding to the moments of time when
the trajectory of the drive system crosses the Poincar\'e cross
section ($x_2=0$) with a positive value of $dx_2/dt$.}
\label{fig7}
\end{figure}

Further decrease of the coupling strength again results in the
onset of synchronization when the value of coupling is in the
second interval of synchronization which takes place between $g
\approx 1.5$ and $g \approx 1.9$ (see Fig.~\ref{fig5}). The synchronized
chaotic attractor which occurs in this case is shown in
Fig.~\ref{fig7}. In the joint phase space of the drive and response
systems this attractor represents a closed ribbon formed after a
``period-doubling" of the synchronized chaotic attractor studied in
the first interval of synchronization, compare Figs.~\ref{fig4}a
and \ref{fig7}a. Due to the period doubling which occurs in the
response system (the driving system remains unchanged), the
synchronization mapping for this attractor becomes a one-to-two
mapping. Depending on initial conditions, the synchronized response
oscillations settle down on the ribbon with one of two different
phases. As a result, the single synchronized attractor in the
drive-response system is presented as two different attractors in
the drive-response-auxiliary system, one in the synchronization
manifold ${\bf y}={\bf z}$ and the other one outside the manifold
(see Fig.~\ref{fig7}b).

In order to examine the synchronous chaotic attractors in more
detail and understand the properties of the corresponding
synchronization mappings, we study the location and stability of
the limiting sets contained in the attractors. These limiting sets
are the unstable periodic orbits (UPOs) embedded in the chaotic
attractor.

\subsection{UPOs and the synchronization mappings}\label{sec2c}

It is well known that phase locking of the unstable periodic orbits
in the coupled chaotic systems plays a significant role in the
understanding of different regimes of chaos synchronization (see,
for example~\cite{Pikovsky97,Zaks99,Parlitz97,RS_PLA96,Park99}). Any
UPO is characterized by the spectrum of multipliers $\mu_i$ or the
corresponding Lyapunov exponents $\Lambda_i=\ln(\mu_i)/T_{UPO}$,
where $T_{UPO}$ is the period of the UPO. In the analysis of
synchronization in directionally-coupled systems, it is always
useful to split the whole spectrum of Lyapunov exponents into two
groups: the first is given by the dynamics of the autonomous drive
system, and the second comes from the conditional dynamics of the
response system. The conditional dynamics corresponds the behavior
induced in the response system by the driving signal.

Since the drive system is a three-dimensional dissipative system
in the present case, the first group of the Lyapunov exponent
spectrum for each UPO contains three exponents: $\Lambda_1 > 0$,
$\Lambda_2 = 0$, and $\Lambda_3 < 0$. These exponents do not depend
upon the coupling strength or any of the other parameters of the response
system. The second group of exponents in the spectrum is due to the
dynamics of the response system. Following~\cite{Pecora90} we
call these exponents the {\em conditional Lyapunov exponents}
$\Lambda^c_i$. If all of the conditional Lyapunov exponents of a UPO have
negative values, we call that orbit a Stable Response (SR) UPO. If at
least one conditional Lyapunov exponent of the UPO is positive, we
call the orbit an Unstable Response (UR) UPO.

\begin{figure}[h]
\begin{center}
\leavevmode
\hbox{%
\epsfxsize=6.1cm
\epsffile{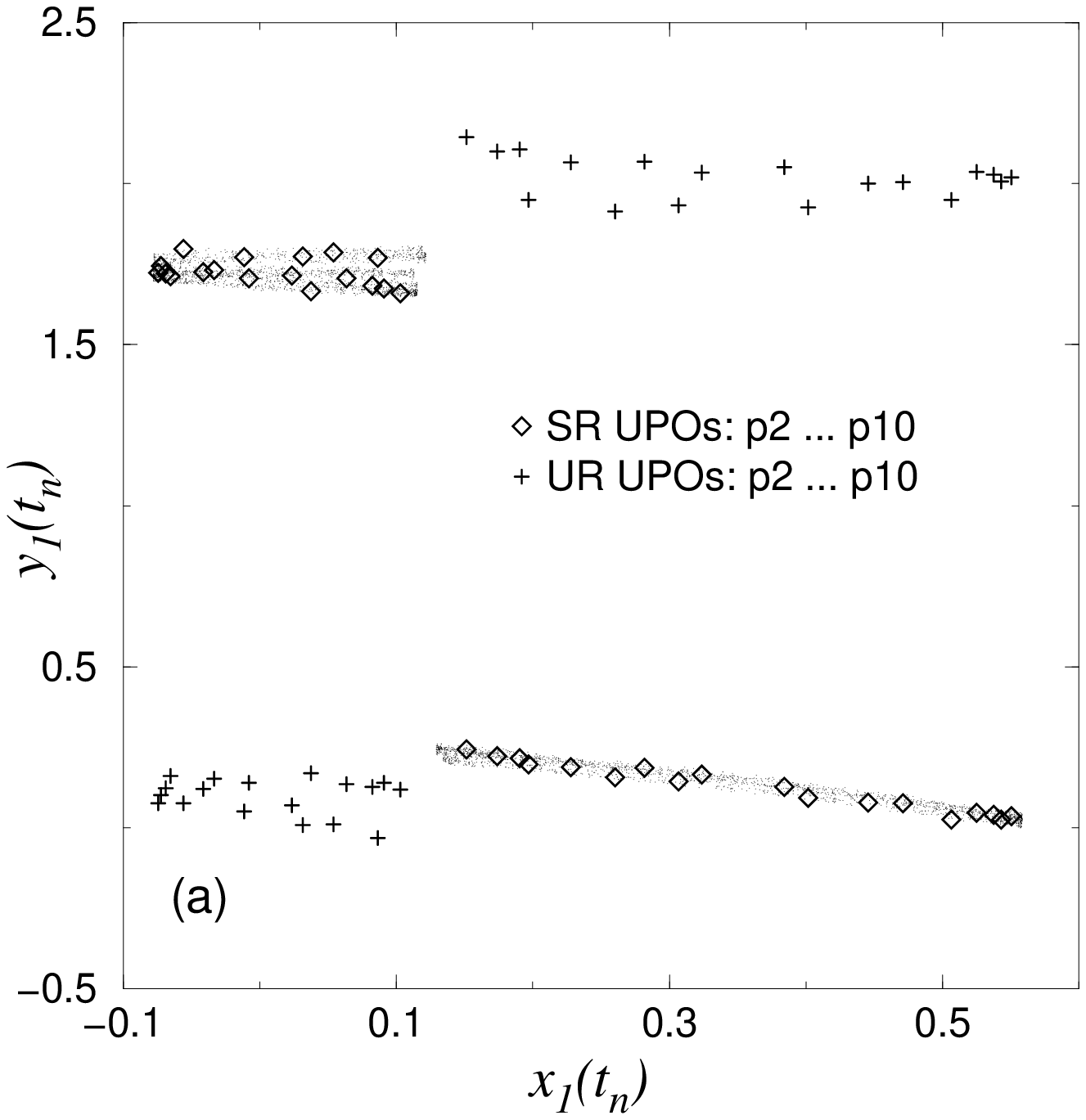}}
\hbox{%
\epsfxsize=6.1cm
\epsffile{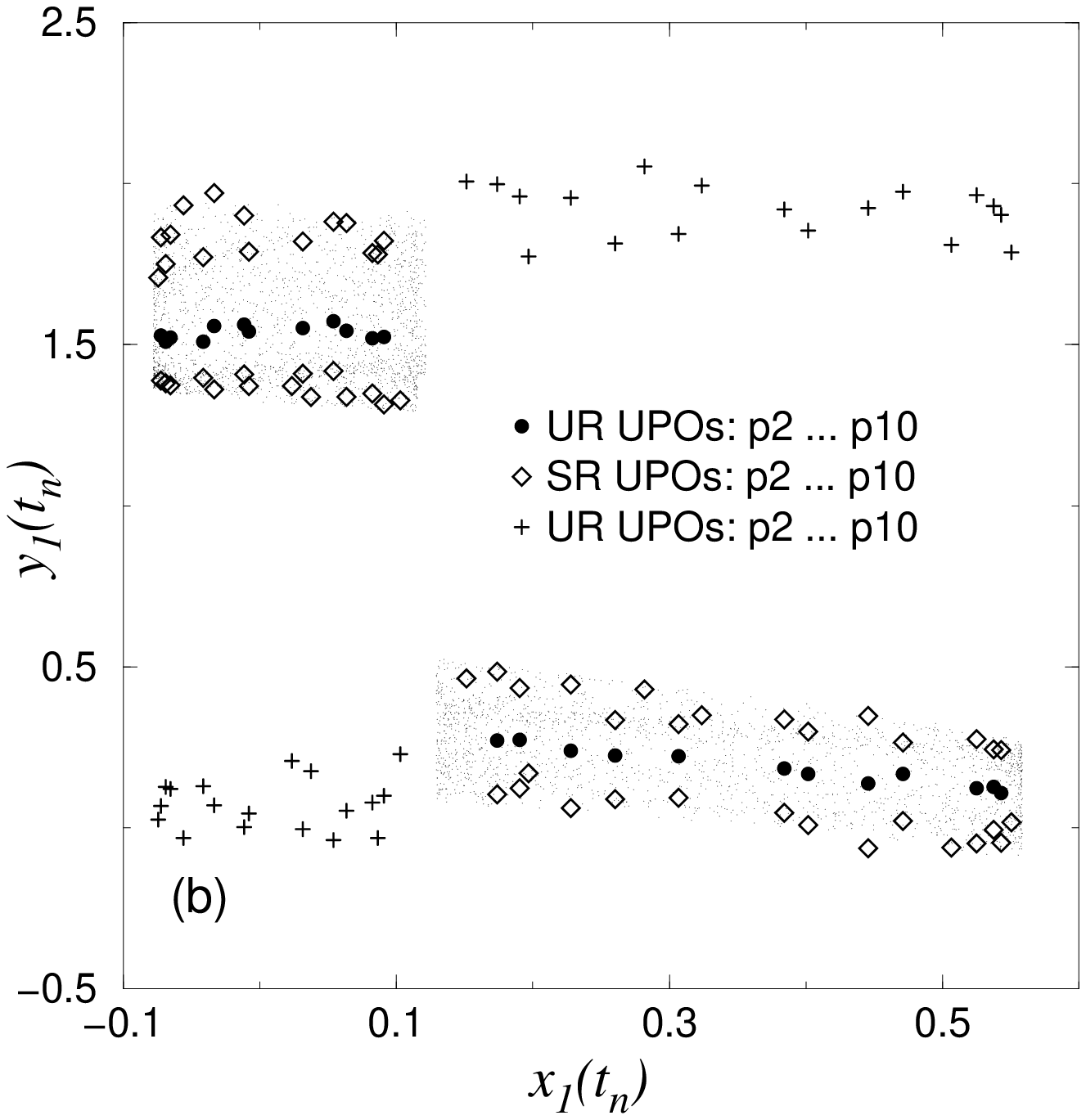}}
\hbox{%
\epsfxsize=6.1cm
\epsffile{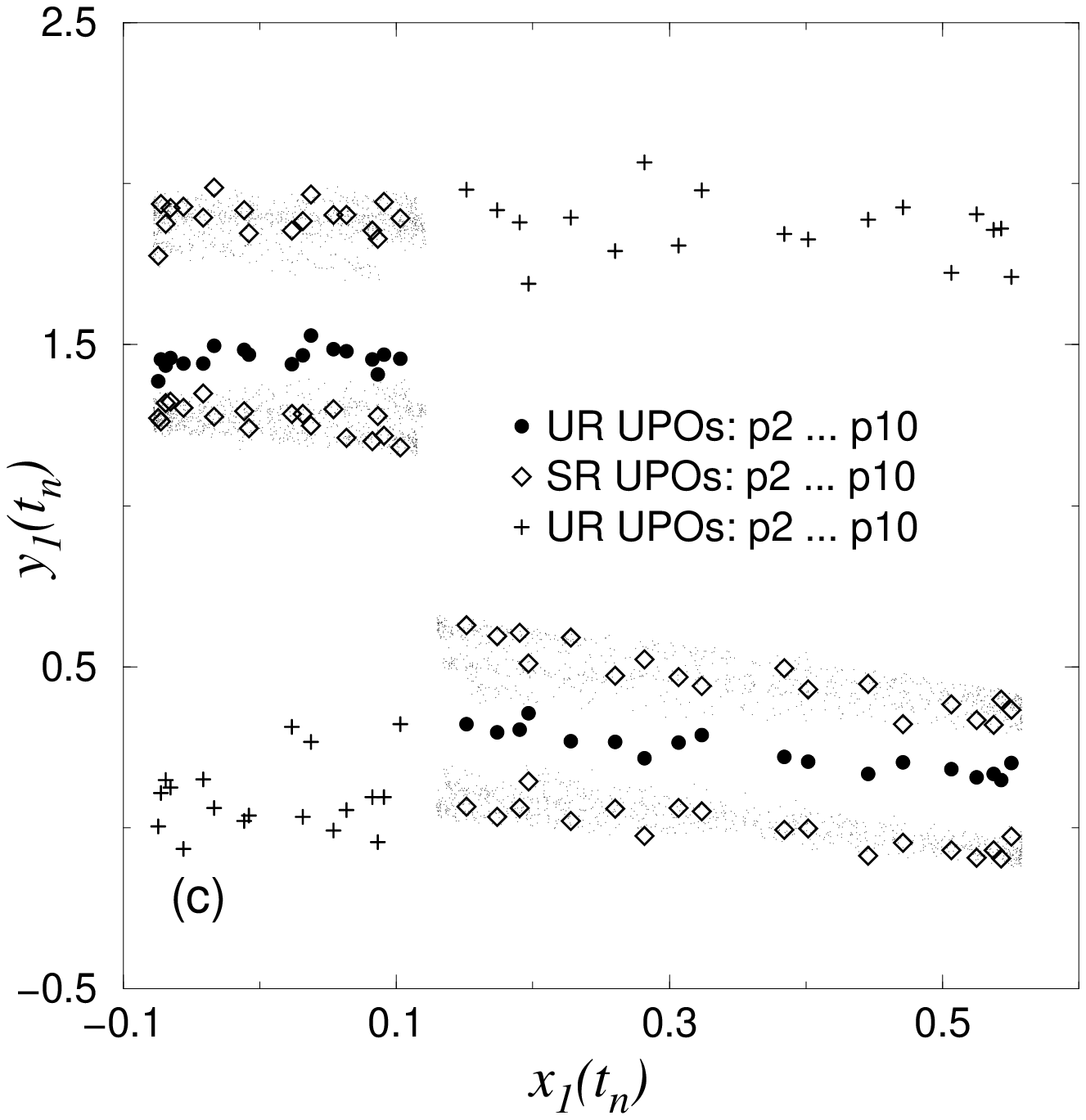}}
\end{center}
\caption{The Poincar\'e cross sections of stable response (SR) UPOs, unstable response (UR)
UPOs and chaotic trajectories in the drive-response system computed
for (a) $g=2.5$, (b) $g=2.0$, and (c) $g=1.8$. }
\label{fig8}
\end{figure}

In our study, we consider all UPOs in the chaotic attractor of the
driving system (shown in Fig.~\ref{fig3}a) up to period six, one of
the period-8 UPOs, and one of the period-10 UPOs. Using the
waveform of $x_1(t)$ of these orbits as a driving force, we
computed the corresponding orbits that are formed in the driven
response system. It is clear that, in this case, the stable
periodic orbits correspond to the SR UPOs and the unstable ones to
the UR UPOs formed in the joint phase space of the drive and
response systems.

The results are shown in Fig.~\ref{fig8}, where we plot the
intersections of the UPOs with the Poincar\'e cross section
($x_2=0$, $dx_2/dt>0$) projected onto the variable plane
$(x_1,y_1)$. In order to see if the UPOs belong to the chaotic
attractor we also plot, in the background, the points on the
Poincar\'e cross section of the chaotic trajectory.

When the systems are synchronized with $g=2.5$, all the SR UPOs are
inside the chaotic attractor, while the UR UPOs do not belong to
the attractor, see Fig.~\ref{fig8}a. The trajectory of the
synchronized chaotic attractor wanders around the SR UPOs which
form the skeleton of the attractor. Since this skeleton does not
contain UR UPOs, the response system always follows the driving
chaotic trajectory in a stable manner. This fact explains the
stability of the identical chaotic oscillations in the response and
auxiliary systems (see Fig~\ref{fig4}b).

Decreasing the value of the coupling parameter $g$ results in a
sequence of bifurcations associated with the SR UPOs. These
bifurcations create a large number of new UPOs, including new UR
UPOs, which are added to the skeleton of the chaotic attractor (see
Fig.~\ref{fig8}b for the case $g=2.0$). Since these new UR UPOs are
inside the chaotic attractor they form regions where the response
behavior for the chaotic driving is unstable. As a result, the
synchronization is destroyed, see Fig.~\ref{fig6}b.

There is the coexistence of two types of UPOs which have different
numbers of unstable directions in the chaotic attractor shown in
Fig.~\ref{fig8}b. This is indicative of the onset of a
non-hyperbolic situation~\cite{Sauer97,Schroer98,Barreto2000}. Note
that even though some kind of non-hyperbolic situation may exist in
the chaotic attractor of the driving system, it does not make any
impact on the response behavior and therefore, makes no impact on
the synchronization between the drive and response systems. The
synchronization is sensitive only to the non-hyperbolicity of the
chaotic attractors contributed by the response system dynamics.
This kind of non-hyperbolic situation is responsible for the
inability to compute the trajectory of the response system if
arbitrarily small perturbations of the system are taken into
account~\cite{Sauer97}. This is the reason why the identical
chaotic oscillations in the response and auxiliary systems do not
follow the same path when even an arbitrary small perturbation are
considered. It also directly indicates the destruction of the
synchronization mapping.

\begin{figure}
\begin{center}
\leavevmode
\hbox{%
\epsfxsize=7.2cm
\epsffile{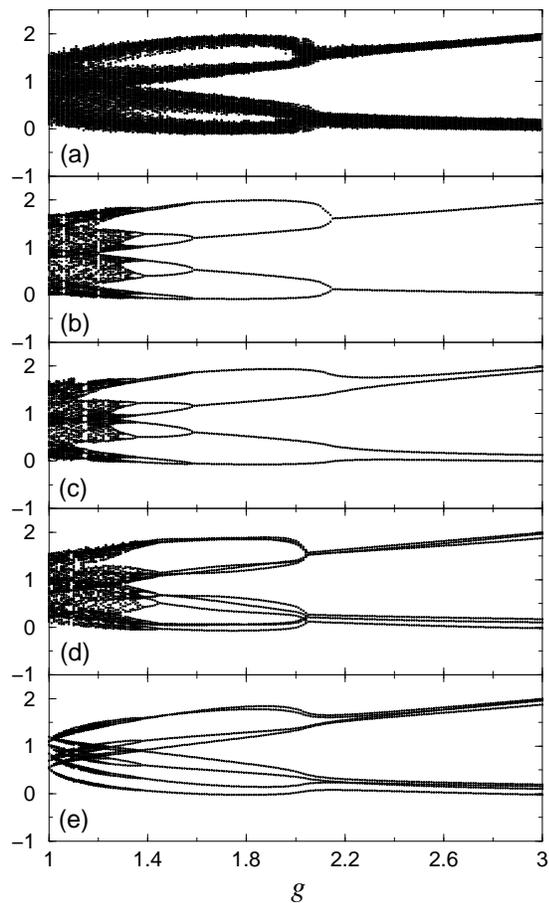}}
\end{center}
\caption{The bifurcation diagram $y_1(t_n)$ vs $g$ computed for the
response system driven by the chaos (a), and by the UPOs: $p2$-(b),
$p4$-(c), $p6$-(d), and $p8$-(e), which are found in the chaotic
attractor of the driving system.}
\label{fig9}
\end{figure}

As one can see from Figs.~\ref{fig6}a and~\ref{fig7}a, the further
decrease of the coupling between the systems results in the period
doubling of the ribbon formed by the chaotic attractor. When this
period doubling ``bifurcation'' is complete, all the SR UPOs and
the chaotic trajectories of the attractor are moved far enough from
the region occupied by the UR UPOs and, as a result, all UR UPOs
again reside only outside the chaotic attractor (see
Fig~\ref{fig8}c). This qualitative change of the chaotic behavior
eliminates the non-hyperbolic situation caused by the dynamics of
the response system and, as a result, the generalized
synchronization is regained (see~\ref{fig7}b). It follows from the
analysis of the SR UPOs that in this regime of chaos
synchronization any point on the UPOs contained in the chaotic
attractor of the driving system maps in to two points located on
the corresponding SR UPOs. Therefore, this synchronization is
characterized by one-to-two synchronization mapping.

In order to understand the properties of this synchronization
mapping we studied the bifurcations of the SR UPOs when the value
of coupling parameter is changed from high to low in the interval
where the ribbon of the chaotic attractor experiences the period
doubling bifurcation. This study, plotted in Fig~\ref{fig9},
reveals a very interesting fact. It turns out that not all the UPOs
experience a period-doubling bifurcation in conjunction with the
bifurcation of the ribbon. Indeed, the orbits which have a period
which is a multiple of four (the period of the chaotic ribbon after
the bifurcation), are not subject to the period-doubling
bifurcations (see Fig~\ref{fig9}c and e). The other UPOs go through
the period doubling bifurcation along with the ribbon.

\begin{figure}
\begin{center}
\leavevmode
\hbox{%
\epsfxsize=7.2cm
\epsffile{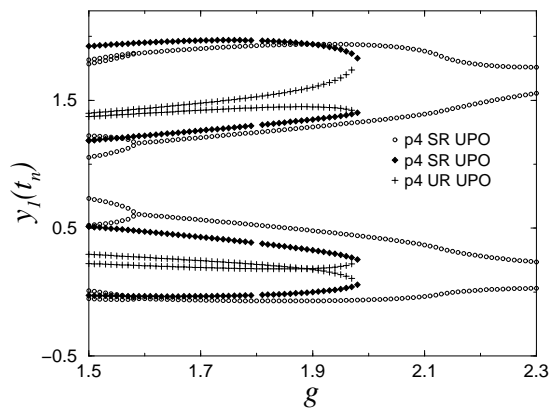}}
\end{center}
\caption{The bifurcation diagram $y_1(t_n)$ vs $g$ computed for the
response system driven by the unstable periodic orbit $p4$ located
in the chaotic attractor of the driving system. The stable response
orbits are indicated by circles and diamonds. The unstable response
orbit is indicated by crosses.}
\label{fig10}
\end{figure}

The difference in the bifurcation scenarios between the
$p(4\times$N) UPOs (with N an integer) and the other UPOs result in
different synchronization mappings occurring when the response
system is driven by these different UPOs. After the period doubling
bifurcation, the orbits whose periods are not multiples of 4 have a
one-to-two synchronization mapping (see Fig.~\ref{fig9}b and d),
while the $p(4\times$N) orbits do not bifurcate and their
synchronization mapping remains one-to-one (see Fig.~\ref{fig9}c
and e). This situation raises the following question: How do these
one-to-one synchronization mappings for the $p(4\times$N) orbits
fit into the one-to-two synchronization mapping which occurs for
all trajectories of the chaotic attractor? The answer to this
question comes from the analysis of the stable response images to
periodic driving with the $p(4\times$N) orbits. When the ribbon
bifurcates, the $p(4\times$N) orbits respond with the creation of a
new pair of response orbits. One is SR UPO and the other is UR UPO
(this bifurcation scenario for the $p4$ orbit is presented in
Fig.~\ref{fig10}). Therefore, despite the fact that periodic
synchronization for $p(4\times$N) is characterized by a one-to-one
mapping, each point on the $p(4\times$N) orbit has two stable
images in the phase space of the response system. Having two SR
UPOs for each $p(4\times$N) orbit, the orbits are able to conform
to the one-to-two chaos synchronization mapping.

This example allows us to draw the following conclusions on the
properties of generalized chaos synchronization with a
multi-valued mapping.
\begin{itemize}
\item The synchronized chaotic attractor represents a ribbon of chaotic
trajectories which does not contain limiting sets with unstable
response dynamics. Each point in the chaotic attractor of the
driving system maps into $M$ points in the chaotic attractor in the
joint phase space of the drive-response system (in the presented example, $M=2$). The value of the number $M$ is invariant over the attractor.
\item These $M$ points are located in the chaotic ribbon in a such way that they represent the different phases of the motion along the ribbon. Any trajectory of the synchronized attractor passes through all phases in sequential order. Therefore, the multi-valued synchronization map can be presented as the cyclic sequence of the single-valued continuous maps.
\item The UPOs of the chaotic attractor in the drive system have SR UPOs, each of which is characterized by the mapping where one point on
the driving UPO maps into $M$ points of the corresponding SR UPO.
The exception to this rule includes only the UPOs of the driving
chaotic attractor with periods that are multiples of the period of
the chaotic ribbon. Each point on these special UPOs can map into a
single point of the corresponding SR UPO. However, these UPOs
should therefore have $M$ stable response UPOs.
\end{itemize}

\section{Functions in the theory of generalized
synchronization of chaos}\label{sec3}

In this section we formulate the theoretical framework for the
synchronization mappings which occur in those regimes of chaos
synchronization similar to the one considered on the previous
section. The main goal here is to link the properties of the
synchronization mappings with the stability characteristics of
the response system behavior.

As it was shown in the previous section, the use of a Poincar\'e cross
section enables one to simplify the analysis of the synchronization
mapping by studying the dynamics of the maps. In a rather general case,
the maps occurring in the drive and response systems can be presented as
maps with a skew product structure of the form:
\begin{eqnarray}
 {\bf x}_{n+1}&=&{\bf F} ({\bf x}_n) \label{mapx} \\
 {\bf y}_{n+1}&=&{\bf G}_k({\bf x}_n,{\bf y}_n)\label{mapy},
\end{eqnarray}
where ${\bf x}$ and ${\bf y}$ are vectors consisting of the variables
in the drive, response and auxiliary systems respectively, and $k$ is a
parameter depending on the strength of the coupling.

Since the first studies of chaos synchronization between {\it
non-identical} systems \cite{AVR86}, synchronization was interpreted
as the system behaving in such a way that for any orbit $({\bf
x}_n,{\bf y}_n)$, the coordinate ${\bf y}_n$ is just a function of
the ${\bf x}_n$, once transients die out. The first rigorous
results on generalized synchronization were obtained for
cases where there exists a smooth manifold belonging to the graph
of a function ${\bf y}=h({\bf x})$ which is stable in the transversal
direction and which contains the chaotic attractor corresponding to
the synchronized oscillations ~\cite{Hunt,Josic}. This approach relies
on the theory of normally hyperbolic invariant manifolds.

In the most of the studies of different regimes of generalized chaos
synchronization, it is observed that in some region of $k$-values, a
function $h$ seems to exist, but it is not smooth. As the result, the
graph of $h$ is a complicated geometrical object (see for example
\cite{Hunt,Pyragas}). In this section we discuss recent rigorous
results on the properties of non-smooth functions $h$ and present
new results for the case of multi-valued mappings.

\subsection{Results and examples concerning single-valued functions $h$}\label{sec3a}

To make the paper self-contained, we first briefly discuss the rigorous
results on the existence and continuity of the synchronization function
$h$ which were obtained in a previous work~\cite{Afr2000}. The paper
demonstrates that the function $h$ is H\"older-continuous if the
contraction rate in the response system is small and
Lipschitz-continuous if it is greater than a critical value.
Assume that ${\bf x}\in\Re^m$  and ${\bf y}\in\Re^n$, and that
${\bf G}_k$ is continuous and ${\bf F}$ is a homeomorphism (i.e.
${\bf F}$ is continuous and invertible with an inverse ${\bf
F}^{-1}$ which is continuous). The dynamics in the joint phase
space of the drive and response systems is determined by a map
$\phi_k:({\bf x}_n,{\bf y}_n)\mapsto ({\bf x}_{n+1},{\bf
y}_{n+1})$. We also assume that in the joint phase space there
exists a basin of attraction $B=B_{xy}\in\Re^{m+n}$, i.e.,
$\phi_k(B)\subset Int(B)$ for any $k\in S$, where $S$ is a region
in $k$-space in which the systems (\ref{mapx}) and (\ref{mapy})
exhibit master-slave chaos synchronization. Without loss of
generality, we assume that $B_{xy}=B_x\times B_y$, i.e., $B_{xy}$ is
a rectangle, where $B_x$ (resp. $B_y$) is a ball in {\bf x}-space
(resp. {\bf y}-space).

Denote by $\Ac_k$ the maximal attractor in $B_{xy}$, i.e.,
$\Ac_k=\cap_{n=0}^{\infty}\phi^n_k(B_{xy})$. Because of our
assumptions,
%%%%%%%%%%%%%%%
\begin{equation}\label{ms}
\lim_{n\rightarrow \infty} | {\bf y}_n-{\bf \tilde y}_n|=0,
\end{equation}
%%%%%%%%%%%%%%%%
where $({\bf x}_n,{\bf y}_n)=\phi_k^n({\bf x}_0,{\bf y}_0)$, $({\bf
x}_n,{\bf \tilde y}_n)=\phi_k^n({\bf x}_0,{\bf \tilde y}_0,$) and
$({\bf x}_0,{\bf y}_0)$, $({\bf x}_0,{\bf \tilde y}_0)$ are
arbitrary points in $B_{xy}$.

To avoid non-essential technicalities it was supposed
in~\cite{Afr2000} that the limit in (\ref{ms}) is achieved
monotonically, i.e.
%%%%%%%%%%%%%%%%%%%%%%%%%%%
\begin{equation}\label{ms1}
|{\bf y}_{n+1}-{\bf \tilde y}_{n+1}|\leq | {\bf y}_n-{\bf
\tilde y}_n|
\end{equation}
%%%%%%%%%%%%%%%%%%%%%%%%%%
for any $k\in S$ and any $n\geq 0$.

Let $\Ac_{k,x}:=\Pi_x\Ac_k$ be the image of $\Ac_k$ under the
natural projection $\Pi_x$ to $\Re^m$.

\bigskip

%%%%%%%%%%%%%%%%%%%%%%%%%%%%%%%%%%%%%%%%%%%%%%%%%%%%%%%%%%%%%%%%%%%%%%%%%%%%
{\bf Theorem 1A}~\cite{Afr2000}. {\it (Existence)\label{existence}
Under assumptions (\ref{ms}) and (\ref{ms1}), there exists a
function $h:\Ac_{k,x}\rightarrow \Re^n$ such that
$\Ac_k=graph(h)$.}

%%%%%%%%%%%%%%%%%%%%%%%%%%%%%%%%%%%%%%%%%%%%%%%%%%%%%%%%%%%%%%%%%%%%%%%%%%%%

\bigskip

We emphasize that it is the inversibility of ${\bf F}$ that
ensures that $h$ is a single-valued function. Continuity plays no role
at this stage.

%%%%%%%%%%%%%%%%%%%%%%%%%%%%%%%%%%%%%%%%%%%%%%%%%%%%%%%%%%%%%%%%%%%%%%%%%%%%%%%%%%
{\bf Theorem 2A}~\cite{Afr2000}. {\it (Continuity) Under the
assumptions of Theorem 1A, the function $h$ is continuous.}

%%%%%%%%%%%%%%%%%%%%%%%%%%%%%%%%%%%%%%%%%%%%%%%%%%%%%%%%%%%%%%%%%%%%%%%%%%%%%%%%%%

\bigskip

To obtain more detailed characteristics of this functional
dependence, additional assumptions were made. It was assumed that
%%%%%%%%%%%%%%%%%%%%%%%%%%%
\begin{equation}\label{ms2}
|{\bf y}_{n+1}-\tilde{{\bf y}}_{n+1}|\leq k_1 | {\bf
y}_n-\tilde{{\bf y}}_n|
\end{equation}
%%%%%%%%%%%%%%%%%%%%%%%%%%
where $k_1<1$. Of course, the parameter $k_1$ is a function of $k$.
For the sake of simplicity, we assume that $k_1=k$. Thus,
%%%%%%%%%%%%%%%%%%%%%%%%%%%
\begin{equation}\label{ms3}
|{\bf y}_{n+1}-\tilde{{\bf y}}_{n+1}|\leq k | {\bf y}_n-\tilde{{\bf
y}}_n|,\quad 0<k<1.
\end{equation}
%%%%%%%%%%%%%%%%%%%%%
It follows that
%%%%%%%%%%%
\begin{equation}\label{lip0}
|{\bf G}_{k}({\bf x},{\bf y})-{\bf G}_{k}({\bf x},\tilde{{\bf
y}})|\leq k |{\bf y}-\tilde{{\bf y}}|
\end{equation}
%%%%%%%%%%%%%
for any $({\bf x},{\bf y})$, $({\bf x},\tilde{{\bf y}})\in B_{xy}$.

Assumption (\ref{ms3}) implies that $| {\bf y}_n-\tilde{{\bf
y}}_n|$ goes to zero exponentially fast, and this fact allows one
to prove that $h$ is H\"older continuous provided that the
functions ${\bf F}$ and ${\bf G_{k}}$ have good smooth properties,
or provided at the least that they are Lipschitz-continuous.

Considering the forward and backward dynamics of the driving
system (\ref{mapx}) we assume the following properties:
%%%%%%%%%%%%%%%%%%
\begin{equation}\label{lip1}
|{\bf F}({\bf x})-{\bf F}(\tilde{{\bf x}})|\leq \gamma_{+} |{\bf
x}-\tilde{{\bf x}}|
\end{equation}
%%%%%%%%%%%%%%%%
and
%%%%%%%%%%%%%%%%
\begin{equation}\label{lip2}
|{\bf F}^{-1}({\bf x})-{\bf F}^{-1}(\tilde{{\bf x}})|\leq
\gamma_{-} |{\bf x}-\tilde{{\bf x}}|\,,
\end{equation}
%%%%%%%%%%%%%%%%
where $\gamma_{-},\gamma_{+}\geq 1$. When Lyapunov exponents do
exist (when the dynamics are differentiable with additional
suitable conditions), the quantities $\log \gamma_{+},\log
\gamma_{-}$ play the role of the forward and backward greatest
Lyapunov exponents respectively, and $\log k$ plays the role of
the conditional Lyapunov exponent.

It is assumed that function ${\bf G}_k({\bf x},{\bf y})$ of the
response system (\ref{mapy}) is Lipschitz continuous with respect
to ${\bf x}$, i.e. for any $({\bf x},{\bf y})$, $(\tilde{{\bf
x}},{\bf y})\in B_{xy}$,
%%%%%%%%%%%%%%%%%%%
\begin{equation}\label{lip3}
|{\bf G}_{k}({\bf x},{\bf y})-{\bf G}_{k}(\tilde{{\bf x}},{\bf
y})|\leq
\eta |{\bf x}-\tilde{{\bf x}}|
\end{equation}
where $\eta>0$.

\bigskip

%%%%%%%%%%%%%%%%%%%%%%%%%%%%%%%%%%%%%%%%%%%%%%%%%%%%%%%%%%%%%%%%%%%%%%%%%%%%%%%%%%%%%%%%%%
{\bf Theorem 3A}~\cite{Afr2000}. {\it (H\"older property) Under
assumptions  (\ref{lip0})-(\ref{lip3}) the function $h$ is H\"older
continuous, i.e. for any $0<\alpha<\alpha_{c}$, and $x,\tilde{x}\in
\Ac_{k,x}$ one has:}
%%%%%%%%%%%%%%
\begin{equation}\label{holder1}
|h({\bf x})-h(\tilde{{\bf x}})|\leq 2\rho |{\bf x}-\tilde{{\bf
x}}|^{\alpha}
\end{equation}
%%%%%%%%%%%%%%%%
{\it where}
%%%%%%%%%%%%
\begin{equation}\label{holder2}\alpha\leq \alpha_c:=
\frac{1}{1-\frac{\log(\gamma_{+}\gamma_{-})}{\log k}}\;,
\end{equation}
%%%%%%%%%%%%
{\it and $\rho\geq \rho_c$, where   $\rho_c$ is the solution of the
equation:}
%%%%%%%%%%%%%%%%%%%%%%%%%%%
$$
\rho=\frac{\eta}{\gamma_{+}-k}\,\rho^{\frac{\log
(\gamma_{+}\gamma_{-})}{\log k}+1}
\exp\Big(\frac{\log k-\log |B_{y}|}{\log k}\log
(\gamma_{+}\gamma_{-})\Big)\,\cdot
$$

%%%%%%%%%%%%%%%%%%%%%%%%%%%%%%%%%%%%%%%%%%%%%%%%%%%%%%%%%%%%%%%%%%%%%%%%%%%%%%
{\bf Theorem 4A}~\cite{Afr2000}. {\it (Lipschitz property) Under
the conditions of Theorem 3A and provided that}
%%%%%%%%%%%
\begin{equation}\label{lip}
0<k<\frac{1}{\gamma_{-}}\;,
\end{equation}
%%%%%%%%%%%%%
{\it the function $h$ is Lipschitz-continuous, i.e.}
%%%%%%%%%%%%%%
\begin{equation}\label{lip4}
|h({\bf x})-h(\tilde{{\bf x}})|\leq L|{\bf x}-\tilde{{\bf x}}|
\end{equation}
%%%%%%%%%%%%%%%%
{\it where} $L\geq
L_c:=\frac{\eta\gamma_{-}}{1-k\gamma_{-}}\,\cdot$

%%%%%%%%%%%%%%%%%%%%%%%%%%%%%%%%%%%%%%%%%%%%%%%%%%%%%%%%%%%%%%%%

The considered theorems give a pretty clear picture of the complexity
of the synchronization mappings which usually occur in different
regimes of generalized synchronization of chaos. It follows from
these results that even in the case of {\em differentiable
generalized synchronization}~\cite{Hunt}, the change of the
parameters toward the border of the synchronization zone will
gradually reduce the rate of contraction in the response system. At
some critical value of the contraction rate (see Theorem 3A and 4A),
the smoothness of synchronization mapping will be destroyed while the
systems remain synchronized with a non-differentiable function.

{\bf Example:} \ \ To confirm the validity of results above, let
us consider coupled H\'enon maps.

We consider the linear coupling of two {\it non-identical}
H\'enon-type maps. For certain values of coupling strength, the
hypothesis of our theorems are fulfilled.

Let $f$ be the following invertible map: $f:[0,1]^{2}\rightarrow
[0,1]^{2}$, $f(x_{1},y_{1})=(x_{1}',y_{1}')$ with
%%%%%%%%%%%%%%%%%%%%%%%%%%%
\begin{eqnarray}
 x_{1}'&=&y_{1}\nonumber \\
 y_{1}'&=&f_{1}(y_{1})+bx_{1} \nonumber
\end{eqnarray}
%%%%%%%%%%%%%%%%%%%%%%%%%%%
where $f_{1}:[0,1]\rightarrow [0,1]$ is Lipschitz, with Lipschitz
constant $\gamma_{1}$, $0<b\leq 1$. Let $f_1$ be such that
$f_1(y_1)\leq \frac{\gamma_1}{2}\; ,\forall y_1\in[0,1]$, and
$(1+b_1)\frac{\gamma_1}{2}\leq 1$. The map $f$ is a homeomorphism
of the unit square.
%%%%%%%%%%%%%%%%%%%%%%%%%%%%%%%%%%%%%%%%%%%%%%%%%%%%%%%%
Let $g_{c}$ be the following map: $g_{c}:[0,1]^{4}\to [0,1]^{2}$,
$g_{c}(x_{1},y_{1},x_{2},y_{2})=(x_{2}',y_{2}')$ with
\begin{eqnarray}
 x_{2}'&=&y_{2}+c(y_{1}-y_{2})\nonumber\\
 y_{2}'&=&f_{2}(y_{2})+b_{2}x_{2}+
 c\big((f_{1}(y_{1})+b_{1}x_{1})-(f_{2}(y_{2})+b_{2}x_{2})\big)
 \nonumber
\end{eqnarray}
%%%%%%%%%%%%%%%%%%%%%%%%%%%%%%%%%%%%%%%%%%%%%%%%%%%%%%%%
where $0\leq c \leq 1$, $f_{2}:[0,1]\rightarrow [0,1]$ is
Lipschitz, with Lipschitz constant $\gamma_{2}$,
$(1+b_2)\frac{\gamma_2}{2}\leq 1$ and $f_2(y_2)\leq
\frac{\gamma_2}{2}\; ,\forall y_2\in[0,1]$.
For convenience, we shall use the following shorthand:
$v_{1}=(x_{1},y_{1})$ and $v_{2}=(x_{2},y_{2})$.

To have contraction in $v_{2}$-space, that is $\vert
g_{c}(v_{1},v_{2})-g_{c}(v_{1},\tilde{v}_{2})\vert
\leq k\vert v_{2}-\tilde{v}_{2}\vert$  with $0<k<1$, the following condition must be
satisfied \footnote{We use the norm $|v|=|x|+|y|$, where
$v=(x,y)$.}:
%%%%%%%%%%%%%%%%%%%%%%%%%%%%%%
$$
c>1-\frac{1}{1+b_{2}+\frac{\gamma_{2}}{2}}\,\cdot
$$
%%%%%%%%%%%%%%%%%%%%%%%%%%%%%%

The forward and backward expansion rate of ${\bf F}$ are
respectively given by: $\gamma_{+}=1+\frac{\gamma_{1}}{2}$ and
$\gamma_{-}=b_{1}^{-1}$.

Theorem 3A holds, i.e. the function $h$ is H\"older continuous and
one can easily check the following expression for the H\"older
exponent,

%%%%%%%%%%%%%%%%%%%%%%%%%%%%%%%%%%%%%%%%%%%%%%%%%%%%%%%%%%%%
$$
\alpha_{0}=
\frac{\log [(1-c)(1+b_2+\frac{\gamma_{2}}{2}]}
{\log\left(
\frac{b_1(1-c)(1+b_2\frac{\gamma_{2}}{2})}{1+\frac{\gamma_{1}}{2}}\right)}\;,
$$

%%%%%%%%%%%%%%%%%%%%%%%%%%%%%%%%%%%%%%%%%%%%%%%%%%%%%%%%%%%%

The condition (\ref{lip}) in theorem 4 reads
%%%%%%%%%%%%%%%%%%%%%%%%%%%%%%%%%%%%%%%%%%%%%
$$
c>1-\frac{b_1}{1+b_2+\frac{\gamma_{2}}{2}}\;,
$$
%%%%%%%%%%%%%%%%%%%%%%%%%%%%%%%%%%%%%%%%%%%%%
According to theorem 4A, this means that the function $h$ is
Lipschitz continuous.

%%%%%%%%%%%%%%%%%%%%%%%%%%%%%%%%%%%%%%%%%%%%%%%%%%%%%%%%%%%%%%%%%%%%%%

\subsection{Multi-valued function $h$}\label{sec3b}

The framework of chaos synchronization discussed above applies only
to the cases when the function $h$ is a single-valued function,
because of the assumption (\ref{ms}). As a result this framework is
not applicable to the case of synchronization presented in
Fig.~\ref{fig7} where  each point on the chaotic attractor
$\Ac_{k,x}$ in driving system maps into two different points of the
synchronized chaotic attractor $\Ac_k$. In this case $B_y$ is not a
simply connected region, and assumption (\ref{ms}) cannot be used.
Below we reformulate the results on single-valued function in order
to extend it to the case of multi-valued function.

Assume that $B_y=\cup_{i=1}^{p} B_y^i$, where $B_y^i$ are pairwise
disjoint closed balls in the ${\bf y}$-space, $B_x$ is a closed
ball in the $x$-space, and $B_{xy}=B_x\times B_y$.

We also assume that monotonic synchronization occurs in $B_{xy}$,
i.e.
%%%%%%%%%%%%%%%
\begin{equation}\label{ms'}
\lim_{n\rightarrow \infty} | {\bf y}_n-{\bf \tilde y}_n|=0,
\end{equation}
%%%%%%%%%%%%%%%%
and
%%%%%%%%%%%%%%%%%%%%%%%%%%%
\begin{equation}\label{ms1'}
|{\bf y}_{n+1}-{\bf \tilde y}_{n+1}|\leq | {\bf y}_n-{\bf
\tilde y}_n|
\end{equation}
%%%%%%%%%%%%%%%%%%%%%%%%%%
where $({\bf x}_n,{\bf y}_n)=\phi_k^n({\bf x}_0,{\bf y}_0),$ $({\bf
x}_n,{\bf \tilde y}_n)=\phi_k^n({\bf x}_0,{\bf \tilde y}_0,$) and
$({\bf x}_0,{\bf y}_0)$, $({\bf x}_0,{\bf \tilde y}_0)$ are
arbitrary points in $B$ such that ${\bf y}_0$ and
${\bf\tilde{y}}_0$ belong to the same ball $B_y^i$ for some $i$.

Denote by $\Ac_k$ the maximal attractor in $B_{xy}$, i.e.
$\Ac_k=\cap_{n=0}^{\infty}\phi^n_k(B_{xy})$ and let
$\Ac_{k,x}:=\Pi_x\Ac_k$ be the image of $\Ac_k$ under the natural
projection $\Pi_x$ to $\Re^m$.

In this situation some extensions of theorems 1-4 hold.

%%%%%%%%%%%%%%%%%%%%%%%%%%%%%%%%%%%%%%%%%%%%%%%%%%%%%%%%%%%%%%%%%%%%%%%%%%%%
{\bf Theorem 1B}~. {\it (Existence)\label{existence} Under
conditions (\ref{ms'}) and (\ref{ms1'}), and provided that for
any ${\bf x}\in\Ac_{k,x}$ and for any $i$, $1\leq i\leq p$,  there
exist ${\bf y}_i\in B_y^i$ such that $({\bf x},{\bf
y}_i)\in\Ac_{k}$, there exist $p$ functions
$h_i:\Ac_{k,x}\rightarrow \Re^n$, $i=1,...,p$ such that
$graph(h_i)\subset B_x\times B_y^i$ and $\Ac_{k}=\cup_{i=1}^p
graph(h_i)$.}

%%%%%%%%%%%%%%%%%%%%%%%%%%%%%%%%%%%%%%%%%%%%%%%%%%%%%%%%%%%%%%%%%%%%%%%%%%%%

{\bf Scheme of the proof.}
%%%%%%%%%%%%%%%%%%%%%%%%%
The proof is mainly the same as the one for single-valued case
(Theorem {\bf 1A}). However we need to emphasize the following
facts. For any ${\bf x}\in \Ac_{k,x}$ and for any $n\geq 0$, the
set ${\bf \phi}_k^n({\bf x},\cup_{i=1}^{p}B_y^i))$ has $p$
connected components inside $({\bf F}^n({\bf
x}),\cup_{i=1}^{p}B_y^i))$ and every set $({\bf F}^n({\bf
x}),B_y^i))$ contains one and only one of them. Indeed, if we
assume that $({\bf F}^n({\bf x}),B_y^i))$ contains more than one
connected components for some $i$ then it will imply that there
exist a $j$ such that $({\bf F}^n({\bf x}),B_y^j))$ contains no
components. Since ${\bf F}$ is one-to-one, it means that there are
no points of the attractor in  $({\bf F}^n({\bf x}),B_y^j))$ and we
have a contradiction. From this point forward, one can follow the
proof of Theorem {\bf 1A}. $\;\Box$

\bigskip

%%%%%%%%%%%%%%%%%%%%%%%%%%%%%%%%%%%%%%%%%%%%%%%%%%%%%%%%%%%%%%%%%%%%%%%%%%%%%%%%%%
{\bf Theorem 2B}~. {\it (Continuity) Under assumptions of Theorem
1B, the functions $h_i$  are continuous, for $i=1,...,p$.}

%%%%%%%%%%%%%%%%%%%%%%%%%%%%%%%%%%%%%%%%%%%%%%%%%%%%%%%%%%%%%%%%%%%%%%%%%%%%%%%%%%

{\bf Scheme of the proof.}
%%%%%%%%%%%%%%%%%%%%%%%%%
Let ${\bf x}$ and $\tilde{{\bf x}}$ be close to each other and consider the points
 $({\bf x},h^i({\bf x}))$ and  $(\tilde{{\bf x}},h^i(\tilde{{\bf x}}))$. We show now that
$\phi_k^{-1}({\bf x},h^i({\bf x}))$ and $\phi_k^{-1}(\tilde{{\bf x}},h^i(\tilde{{\bf x}}))$
belong to the sets $\big({\bf F}^{-1}({\bf x}),B_y^j\big)$ and
$\big({\bf F}^{-1}(\tilde{{\bf x}}),B_y^j)\big)$, respectively, with the same number $j$.

Assuming the contrary, i.e.,  $\phi_k^{-1}({\bf x},h^i({\bf x}))\in\big({\bf F}^{-1}({\bf x}),B_y^j\big)$
but
$\phi_k^{-1}(\tilde{{\bf x}},h^i(\tilde{{\bf x}}))\in\big({\bf F}^{-1}(\tilde{{\bf x}}),B_y^{j'})\big)$
and $j'\neq j$.
Therefore $\phi_k\big({\bf F}^{-1}(\tilde{{\bf x}}),B_y^{j'})\big)$ belongs to
$(\tilde{{\bf x}},,B_y^i)$ and $\phi_k\big({\bf F}^{-1}(\tilde{{\bf x}}),B_y^{j})\big)$ belongs to
$(\tilde{{\bf x}},,B_y^{i'})$ where $i'\neq i$.
Consider now the points $\big({\bf F}^{-1}({\bf x}),{\bf y}\big)$ and $\big({\bf F}^{-1}(\tilde{{\bf x}}),{\bf y}\big)$,
${\bf y}\in B_y^i$. It follows that $dist\Big(\phi_k\big({\bf F}^{-1}({\bf x}),{\bf y}\big),\phi_k
\big({\bf F}^{-1}(\tilde{{\bf x}}),{\bf y}\big)\Big)$ is bounded from zero provided that
${\bf x}$ and $\tilde{{\bf x}}$ are close enough. Roughly speaking, this distance is greater
than $\frac{1}{2}dist(B_y^j,B_y^{j'})$.
On the other side, we know that
$|{\bf F}^{-1}({\bf x})-{\bf F}^{-1}(\tilde{{\bf x}})|$ is small, therefore by continuity of $\phi_k$,
$dist\Big(\phi_k\big({\bf F}^{-1}({\bf x}),{\bf y}\big),\phi_k\big({\bf F}^{-1}(\tilde{{\bf x}}),{\bf y}\big)\Big)$ is
also small, resulting in a contradiction.

The rest of the proof is the same as in Theorem {\bf 2A}. $\;\Box$

\bigskip
We assume now that conditions similar to (\ref{ms2})-(\ref{lip3})
hold in $B_{xy}$, i.e., inequalities (\ref{ms2})-(\ref{lip0}) hold
provided that ${\bf y}={\bf y}_0$, ${\bf \tilde{y}}={\bf
\tilde{y}}_0$ belong to
the same connected component in the set $B_y$; other conditions
remain the same.

Taking into account the schemes of the proofs of Theorems 1B and 2B, one
can check that the proofs of the following two theorems are
similar to the proofs of Theorems 3A and 4A.

\bigskip

%%%%%%%%%%%%%%%%%%%%%%%%%%%%%%%%%%%%%%%%%%%%%%%%%%%%%%%%%%%%%%%%%%%%%%%%%%%%%%%%%%%%%%%%%%
{\bf Theorem 3B}~. {\it (H\"older property) If the conditions are
satisfied, under assumptions of Theorem 1B, the functions $h_i$,
$i=1,...,p$ are  H\"older continuous, i.e. for any
$0<\alpha<\alpha_{c}$, ${\bf x},\tilde{{\bf x}}\in
\Ac_{k,x}$ one has:}
%%%%%%%%%%%%%%
\begin{equation}\label{holder1'}
|h_i({\bf x})-h_i(\tilde{{\bf x}})|\leq 2\rho |{\bf x}-\tilde{{\bf
x}}|^{\alpha}
\end{equation}
%%%%%%%%%%%%%%%%
{\it where}
%%%%%%%%%%%%
\begin{equation}\label{holder2'}\alpha\leq \alpha_c:=
\frac{1}{1-\frac{\log(\gamma_{+}\gamma_{-})}{\log k}}\;,
\end{equation}
%%%%%%%%%%%%
{\it and $\rho\geq \rho_c$, where $\rho_c$ is the solution of the
equation:}
%%%%%%%%%%%%%%%%%%%%%%%%%%%
\begin{equation}
 \rho=\frac{\eta}{\gamma_{+}-k}\,\rho^{\frac{\log
(\gamma_{+}\gamma_{-})}{\log k}+1} \times
\exp(\Delta(k,\gamma_{+},\gamma_{-}))\, ,
\end{equation}
where
\begin{equation}
\Delta(k,\gamma_{+},\gamma_{-}):= \frac{\log k-\log \big(\max_i|B_{y}^i|\big)}{\log k}\log
(\gamma_{+}\gamma_{-})\,\cdot
\end{equation}

%%%%%%%%%%%%%%%%%%%%%%%%%%%%%%%%%%%%%%%%%%%%%%%%%%%%%%%%%%%%%%%%%%%%%%%%%%%%%%
{\bf Theorem 4B}~. {\it (Lipschitz property) Under conditions of
Theorem 3B and provided that}
%%%%%%%%%%%
\begin{equation}\label{lip'}
0<k<\frac{1}{\gamma_{-}}\;,
\end{equation}
%%%%%%%%%%%%%
{\it the functions $h_i$, $i=1,...,p$, are Lipschitz-continuous,
i.e.}
%%%%%%%%%%%%%%
\begin{equation}\label{lip4'}
|h_i({\bf x})-h_i(\tilde{{\bf x}})|\leq L|{\bf x}-\tilde{{\bf x}}|
\end{equation}
%%%%%%%%%%%%%%%%
{\it where} $L\geq
L_c:=\frac{\eta\gamma_{-}}{1-k\gamma_{-}}\,\cdot$

%%%%%%%%%%%%%%%%%%%%%%%%%%%%%%%%%%%%%%%%%%%%%%%%%%%%%%%%%%%%%%%%

In order to deal with problems in the case of multi-valued function
$h$ we make use of the auxiliary systems
approach~\cite{Abarbanel96}. Consider an auxiliary system which is
a replica of the response system (\ref{mapy}) and given by the
equation of the form
\begin{eqnarray}
{\bf z}_{n+1}&=&{\bf G}_k({\bf x}_n,{\bf z}_n)\,\label{mapz}
\end{eqnarray}
Note that the auxiliary system (\ref{mapz}) can serve, as well as
in the case of standard  generalized synchronization, to indicate
the validity of synchronization.

Let the dynamics in the joint phase space of the drive, response
and auxiliary systems be determined by a map $\psi_k:({\bf
x}_n,{\bf y}_n,{\bf z}_n)\mapsto ({\bf x}_{n+1},{\bf y}_{n+1},{\bf
z}_{n+1})$. Due to the identity between the response and auxiliary
system there always exist an invariant manifold $M_S:{\bf y=z}$.
Let us relate the ball $B_y^i$ to $B_z^i$: $z_0\in B_z^i$ iff
$y_0=z_0\in B_y^i$. An indicator of multi-valued synchronization We
can use the equality
%%%%%%%%%%%%%%%
\begin{equation}\label{mv}
\lim_{n\rightarrow \infty} | {\bf y}_n-{\bf z}_n|=0,
\end{equation}
%%%%%%%%%%%%%%%%
where $({\bf x}_n,{\bf y}_n,{\bf z}_n)=\psi_k^n({\bf x}_0,{\bf
y}_0,{\bf z}_0)$ and $({\bf x}_0,{\bf y}_0,{\bf z}_0)$, is an
arbitrary point in $B$ such that $y_0\in B_y^i$ and $z_0\in B_z^i$
with the same index $i$, as an indicator of multi-valued
synchronization. Indeed, it is simple to see that the following statement holds.

{\bf Theorem 5B}~. Under the assumptions of Theorem 1B the equality
(\ref{mv}) is satisfied provided that  $y_0\in B_y^i$ and $z_0\in
B_z^i$ with the same index $i$.

\bigskip

The behavior of orbits in the attractor can be different. It could
be related to a cyclic repetition, $h_1\rightarrow h_2 \rightarrow
h_1\rightarrow h_2\rightarrow ...$, or a more complex sequence of
$h_i$. It depends on the partition of $\Ac_{k,x}$ into connected
components.

In the simplest case when  $\Ac_{k,x}$ is connected, every orbit in
$\Ac_{k}$ behaves in the same way: for any  $i$ there is a $j$ such
that $\phi_k\big(graph(h_i)\big)=graph(h_j)$, $1\leq i,j\leq p$.
This fact is a direct corollary of Theorem 2B. Indeed, for any two
points $Q_1$ and $Q_2$ in $graph(h_i)$ and any $\epsilon>0$ there
is a collection of points $P_0,...P_N\in graph(h_i)$ such that
$P_0=Q_1$, $P_N=Q_2$ and $dist(P_i,P_{i+1})\leq \epsilon$, $0\leq
i\leq N-1$. Because of the continuity of $\phi_k$, the points
$\phi_k(P_i)$ and $\phi_k(P_{i+1})$ belong to the same branch, say
$graph(h_j)$ for any $i=0,...N-1$. In fact, a trajectory switches
 the disjoint balls in the $y$-space in a particular order, and this order
is uniquely determined by a permutation $i\rightarrow j$.

If $\Ac_{k,x}$ contains more than one connected components then the behavior
of orbits inside each of them determines the same itinerary among branches
$graph(h_i)$. For example, if such a component contains the projection of a periodic point
then the itinerary will be also periodic. But if a component does not contain a periodic orbit
then the itinerary could be non-periodic; in this case $\Ac_{k,x}$ should have infinitely many
connected components.

\section{Conclusions}

The example of generalized synchronization of chaos considered in
Section~\ref{sec2} enables us to explore properties of a new type of
chaos synchronization mapping in which the mapping is a multi-valued function.
Based on analysis of the conditional stability of the synchronous
chaotic behavior of the response system, we come to conclusions concerning
the existence and continuity of the synchronization mappings formed
in these regimes of generalized synchronization. We used the auxiliary
system method to detect the onset of conditional stability both in
the case of single-valued mappings and in the case of multi-valued mappings.

Changing the strength of the coupling between the drive and
response systems, we followed the transition from a regime of
synchronization with single-valued function to a regime with
a double-valued function. These two regimes of synchronization are
separated by a regime of asynchronous chaotic oscillations. Our
numerical analysis showed that this regime of asynchronous
oscillations occupies an interval of the coupling parameter values
where all of the necessary bifurcations occur which are required
for the formation of the new type of synchronization mapping. Analysis of the unstable
periodic orbits (UPOs) contained in the chaotic attractor formed in
the joint phase space of drive and response systems revealed a very
interesting element of this bifurcation scenario. We have found
that not all UPOs in the attractor experience the period doubling
bifurcations as the ribbon containing the chaotic trajectories of
the attractor doubles its period. The UPOs with periods that are multiples
to the period of the post-bifurcation chaotic ribbon remain
qualitatively unchanged. In this case the response system generates
additional UPOs with identical periods via
tangential bifurcations. Thanks to these additional UPOs the newly
formed double-valued synchronization function applies universally to
all orbits of the driving chaotic attractor.

The introduction of multi-valued synchronization mapping into the
concept of the generalized synchronization of chaos provides a new
theoretical framework which is crucial for understanding of chaos
synchronization phenomenon in many cases of synchronization in
directionally coupled systems. Such cases include synchronization
with frequency ratio other then one-to-one where the formation of
multi-valued synchronization mappings is a quite natural occurrence.

\section{Acknowledgments}
The authors are grateful to H.D.I. Abarbanel, K. Josic, L. Kocarev,
L. Pecora and U. Parlitz for helpful discussions. N.F. Rulkov was
sponsored in part by U.S. Department of Energy (grant
DE-FG03-95ER14516) and the U.S. Army Research Office (MURI grant
DAAG55-98-1-0269).

\end{document}